\documentclass[aps,amsmath]{revtex4}
\usepackage{graphicx}

        \newcommand{\be}{\begin{equation}}
        \newcommand{\ee}{\end{equation}}
        \newcommand{\bse}{\begin{subequations}}
        \newcommand{\ese}{\end{subequations}}
        \newcommand{\bea}{\begin{eqnarray}}
        \newcommand{\eea}{\end{eqnarray}}
        \newcommand{\nn}{\nonumber}
        \newcommand{\ban}{\begin{eqnarray*}}
        \newcommand{\ean}{\end{eqnarray*}}
        \newcommand{\half}{\frac{1}{2}}

\newcommand{\abs}[1]{\left| #1\right|}

\renewcommand{\Im}{\mbox{Im}}
\renewcommand{\Re}{\mbox{Re}}

\renewcommand{\d}{{\rm d}}

\begin{document}

\preprint{}

\title{Self-consistent calculations of spectral densities in
the $O(N)$ model: improving the Hartree-Fock approximation by including
nonzero decay widths}

\author{Dirk R\"oder}
\email{roeder@th.physik.uni-frankfurt.de}
\affiliation{Institut f\"ur Theoretische Physik,
Johann Wolfgang Goethe-Universit\"at, \\
Max von Laue-Str.\ 1, D-60438 Frankfurt/Main, Germany}
\author{J\"org Ruppert}
\email{ruppert@phy.duke.edu}
\affiliation{Department of Physics, Duke University,
Box 90305, Durham, NC 27708, USA}
\author{Dirk H.\ Rischke}
\email{drischke@th.physik.uni-frankfurt.de}
\affiliation{Institut f\"ur Theoretische Physik
and Frankfurt Institute for Advanced Studies,\\
Johann Wolfgang Goethe-Universit\"at,
Max von Laue-Str.\ 1, D-60438 Frankfurt/Main, Germany}

\begin{abstract}
We study the linear sigma model with $O(N)$ symmetry 
at nonzero temperature in the framework of 
the Cornwall-Jackiw-Tomboulis formalism.
Extending the set of two-particle irreducible diagrams 
by adding sunset diagrams to the usual Hartree-Fock contributions,
we derive a new approximation scheme which extends the standard Hartree-Fock
approximation by the inclusion of nonzero decay widths. In this approximation, 
in-medium modifications of the meson masses as
well as decay and collisional broadening effects are 
self-consistently taken into account. As compared to the standard 
Hartree-Fock approximation, the temperature for the chiral
symmetry restoring phase transition is lowered by $\sim 20-25 \%$.
For large temperatures, the spectral densities of the
$\sigma$-meson and the pion become degenerate, as required for the 
chirally symmetric phase.
\end{abstract}

\date{\today}
\pacs{11.10.Wx, 12.38.Lg, 12.40.Yx}
\maketitle


\section{Introduction}

For zero quark masses, the QCD Lagrangian exhibits a global chiral 
$U(N_f)_r \times U(N_f)_\ell$ or, equivalently,
$U(N_f)_V \times U(N_f)_A$ symmetry, where $N_f$ is the number of 
quark flavors, and $V\equiv r + \ell, A \equiv r - \ell$. 
In the vacuum, the $U(1)_A$ anomaly \cite{'tHooft:1986nc} breaks 
this symmetry explicitly to $U(N_f)_V \times SU(N_f)_A$.
Furthermore, a nonvanishing 
chiral condensate $\langle \bar{q} q \rangle \sim (300~{\rm MeV})^3$
breaks this symmetry 
spontaneously to $U(N_f)_V$ \cite{Vafa:1984tf},
leading to $N_f^2-1$ pseudoscalar 
Goldstone bosons which constitute
the effective degrees of freedom at low energies.
In nature, chiral symmetry is also explicitly broken by
nonvanishing quark masses, which accounts for the physical masses
of the pions, kaons, etc.

At temperatures of the order of $\sim \langle \bar{q} q \rangle^{1/3}$,
the thermal excitation energy is large enough to expect
the restoration of chiral symmetry.
At such energy scales, the QCD coupling constant is still large,
rendering perturbative calculations unreliable. Thus, one has to resort to
nonperturbative methods to study chiral symmetry restoration.
A first-principle approach is lattice QCD
\cite{Karsch:2001cy}. Lattice QCD calculations have determined the
temperature $T_c$ for chiral symmetry restoration to be of order 150 MeV
at zero quark chemical potential \cite{Laermann:2003cv}. These 
calculations, however, face several technical problems.

The first is that they become numerically very difficult for 
physically realistic, i.e., small, values of the up- and down-quark masses.
Although progress in this direction has been made \cite{Fodor:2004nz},
most studies use unphysically large values.
Another problem is that, at nonzero
chemical potential, lattice QCD calculations are hampered
by the fermion sign problem and
become increasingly unreliable for chemical potentials 
larger than (a factor $\pi$ times) the temperature \cite{deForcrand:2002ci}.

An alternative nonperturbative approach to study chiral symmetry restoration
is via chiral effective theories. These theories have
the same global $U(N_f)_r \times U(N_f)_\ell$ symmetry as QCD
but, since quark and gluons are integrated out, 
do not possess the local $SU(3)_c$ color symmetry of QCD.
The effective low-energy degrees of freedom are 
the (pseudo-) Goldstone bosons of the QCD vacuum, i.e.,
the pseudoscalar mesons. However, in the chirally symmetric phase
these particles become degenerate
with their chiral partners, the scalar mesons. 
Therefore, an appropriate effective 
theory to study chiral symmetry restoration in QCD is the linear sigma model
\cite{Levy,Gell-Mann:1960np}
which treats both scalar and pseudoscalar degrees of freedom on the
same footing.
The advantage of chiral effective theories over lattice QCD
calculations is that their numerical treatment 
(within some many-body approximation scheme) is comparatively simple
and that there is no
problem to consider arbitrary quark chemical potential.

At nonzero temperature $T$, ordinary perturbation theory in terms of 
the coupling constant $g$ breaks down in the sense that one
can no longer order different contributions
according to powers of $g$ \cite{Dolan:1974qd}. 
This is because the new energy scale introduced by the 
temperature can conspire with the typical momentum scale $p$
of a process so that $gT/p$ is no longer of order $g$, but can be of
order $1$ \cite{Braaten:1989kk,Braaten:1989mz}.
Consequently, all terms of order $gT/p$ have to be taken into account
which requires the resummation of certain classes of diagrams.

A convenient resummation method is provided by the 
extension of the Cornwall-Jackiw-Tomboulis (CJT) formalism 
\cite{Cornwall:1974vz} to nonzero temperatures and chemical potentials.
The CJT formalism is equivalent to the $\Phi$-functional 
approach of Luttinger and
Ward \cite{Luttinger:1960ua} and Baym \cite{Baym:1962sx}.
It generalizes the concept of the 
effective action $\Gamma[\bar{\phi}]$ for the expectation value
$\bar{\phi}$ of the one-point function in the presence of
external sources to that for the effective action
\begin{equation} \label{effact}
\Gamma[\bar{\phi},\bar{G}] = S[\bar{\phi}] + \frac{1}{2} \,
{\rm Tr} \ln \bar{G}^{-1}
+ \frac{1}{2} {\rm Tr} ( G^{-1} \bar{G} - 1) + \Gamma_2[\bar{\phi}, \bar{G}]
\end{equation}
for $\bar{\phi}$ and the expectation value $\bar{G}$ of the
two-point function in the presence of external sources. Here, 
$S[\bar{\phi}]$ is the tree-level action, $G^{-1}$ the inverse tree-level
two-point function, and $\Gamma_2[\bar{\phi},\bar{G}]$
the sum of all two-particle irreducible
(2PI) vacuum diagrams with internal lines given by $\bar{G}$. 
(For an extension of this approach
to three- and more-point functions, see
\cite{Norton:1974bm,Kleinert:1982ki,Carrington:2004sn,Berges:2004pu}.)
The stationary points of this functional,
\begin{equation} \label{stat}
\left.\frac{\delta \Gamma[\bar{\phi},\bar{G}]}{\delta \bar{\phi}}
\right|_{\bar{\phi}=\varphi,\bar{G} = {\cal G}} = 0\;, \;\;\;\;
\left.\frac{\delta \Gamma[\bar{\phi},\bar{G}]}{\delta \bar{G}}
\right|_{\bar{\phi}=\varphi,\bar{G} = {\cal G}} = 0\;,
\end{equation}
provide self-consistent
equations for the expectation values of the one- and two-point functions 
$\bar{\phi}$ and $\bar{G}$ in the {\em absence\/} of external sources,
denoted as $\varphi$ and ${\cal G}$, respectively. The
latter is the Dyson-Schwinger equation,
\begin{equation}
{\cal G}^{-1} = G^{-1} + \Pi\;,
\end{equation}
where
\begin{equation} \label{Pi}
\Pi \equiv - 2 \left. \frac{\delta \Gamma_2 [\bar{\phi},\bar{G}]}{\delta
\bar{G}} \right|_{\bar{\phi}= \varphi, \bar{G} = {\cal G}}
\end{equation}
is the self-energy. As long as $\Gamma_2$ contains {\em all\/}
2PI diagrams, the CJT effective action is exact. Of course, it is
practically impossible to compute all 2PI diagrams, and one has
to truncate $\Gamma_2$, for instance at some order in the number of loops.
The advantage of the CJT formalism is that {\em any\/} truncation of
$\Gamma_2$ yields a many-body approximation
scheme which preserves the symmetries of the
tree-level action, if the expectation values $\bar{\phi}$ and $\bar{G}$ 
transform as first- and second-rank tensors of these
symmetries \cite{Baym:1962sx}.
The solution of Eqs.\ (\ref{stat}) is thermodynamically consistent and
conserves the Noether currents.
However, Ward-Takahashi identities for higher-order vertex functions 
may be violated because of the truncation of $\Gamma_2$ 
\cite{vanHees:2002bv}. 
 
The most popular among these many-body approximation
schemes is the Hartree-Fock approximation.
In this case, $\Gamma_2$ solely contains diagrams of so-called
``double-bubble'' topology, cf.\ Fig.\ \ref{paper1} a--c. 
Note that, in Refs.\ 
\cite{Petropoulos:1998gt,Lenaghan:1999si,Lenaghan:2000ey,Roder:2003uz}, this
scheme has been termed ``Hartree'' approximation.
However, since the exchange contributions with respect to internal
indices are in fact included, it is more
appropriate to call it ``Hartree-Fock'' approximation. Neglecting
the exchange contributions leads to the actual Hartree approximation
which, in the case of the $O(N)$ model, has also been termed 
``large-$N$'' approximation \cite{Petropoulos:1998gt,Lenaghan:1999si}.
In the Hartree-Fock approximation, 
the self-energies, which according to Eq.\ (\ref{Pi})
are obtained by cutting lines in the diagrams for $\Gamma_2$,
only consist of ``tadpole'' diagrams, cf.\ Figs.\ \ref{selfenergy_sigma} 
b,c, and \ref{selfenergy_pion} a,c.
For the chiral effective theories, such as the $U(N_f)_r \times
U(N_f)_\ell$ and $O(N)$ models, this approximation scheme
has been studied in great detail 
\cite{Baym:1977qb,Bochkarev:1995gi,Roh:1996ek,Amelino-Camelia:1997dd,Petropoulos:1998gt,Lenaghan:1999si,Lenaghan:2000ey,Roder:2003uz}.
The Hartree-Fock approximation is a very simple approximation scheme,
since tadpole self-energies do not have an imaginary part and,
consequently, all particles are stable quasiparticles. Moreover, since
tadpoles are independent of energy and momentum,
the Dyson-Schwinger equations for the propagators reduce to 
fix-point equations for the in-medium masses.

There are, however, several problems with the Hartree-Fock approximation
related to the truncation of $\Gamma_2$.
For instance, in the case of the $O(N)$ model,
it does not correctly reproduce the order of the chiral
symmetry restoring phase transition 
and it violates Goldstone's theorem in the sense
that the Goldstone bosons do not remain massless for nonzero temperatures
$0 <T< T_c$.
Several ways to remedy this shortcoming have been suggested.
The simplest one is to neglect subleading contributions in $1/N$,
leading to the Hartree (or large-$N$, see discussion above) approximation 
\cite{Petropoulos:1998gt,Lenaghan:1999si}.
Here, the Goldstone bosons remain massless for $T< T_c$
and the transition is of second order, as expected from universality
class arguments \cite{Pisarski:1984ms}.
Another possibility to restore Goldstone's theorem is via
so-called ``external'' propagators 
\cite{Aouissat:1997nu,vanHees:2001ik,VanHees:2001pf,vanHees:2002bv,Aarts:2002dj}. 
These objects are obtained from the Hartree-Fock propagators
by additionally resumming all diagrams pertaining to the
Random Phase Approximation (RPA), with internal lines given
by the Hartree-Fock propagators.
Another way to restore the second-order nature of the chiral 
phase transition is to include so-called ``two-particle point-irreducible'' 
(2PPI) contributions to $\Gamma_2$ 
\cite{Verschelde:1992bs,Verschelde:2000dz,Baacke:2002pi}.

In this work, we are not concerned with these formal shortcomings
of the Hartree-Fock approximation: we focus exclusively on
the case realized in nature where chiral symmetry is
already explicitly broken by (small) nonzero quark masses, such that
the pion is no longer a Goldstone boson and assumes a mass 
$M_\pi \simeq 139.5~{\rm MeV}$ in vacuum \cite{PDBook}.
Our goal in this work is to include the nonzero decay width
of the particles in a self-consistent many-body approximation
scheme. To this end, one has to go beyond the Hartree-Fock
approximation and add other diagrams to $\Gamma_2$ which, upon cutting
lines according to Eq.\ (\ref{Pi}), yield self-energies with
nonzero imaginary part. The most simple of such diagrams, and the
ones considered in the following,
are those of the so-called ``sunset'' topology, cf.\ Fig.\ \ref{paper1} d,e,
leading to the self-energy diagrams shown in Figs.\ \ref{selfenergy_sigma} d,e
and \ref{selfenergy_pion} d. 

The self-energies arising from the sunset diagrams in $\Gamma_2$
depend on the energy and the momentum of the incoming particle.
Therefore, the Dyson-Schwinger equations no longer reduce to
fix-point equations for the in-medium masses. Since
the self-energies now have a nonzero imaginary part, implying
a finite lifetime of the corresponding particles, the spectral densities are
no longer delta-functions. It is therefore convenient to rewrite
the Dyson-Schwinger equations for the propagators into 
a set of self-consistent integral equations for the spectral densities 
which has to be solved numerically.

The paper is organized as follows. In Sec.\ \ref{sectionII} we discuss the
application of the CJT formalism to the $O(N)$ model. In Sec.\ \ref{III}
we discuss our new approximation scheme which extends the standard
Hartree-Fock approximation by including nonzero decay widths. 
Numerical results for the in-medium
masses and decay widths of $\sigma$-mesons and pions, as well
as their spectral densities are discussed in Sec.\ \ref{results}.
Section \ref{conclusion} concludes this paper with a summary of our results.
Technical details are deferred to the appendices.

We denote 4-vectors by capital letters, $X^\mu \equiv(x_0,{\bf x})$, with
${\bf x}$ being a 3-vector of modulus
$x\equiv |{\bf x}|$. 
We use the imaginary-time formalism to compute quantities at
nonzero temperature. Integrals over 4-momentum $K^\mu = (k_0, {\bf k})$ are
denoted as
\be
\int_K \, f(K) \equiv T \sum_{n=-\infty}^{\infty}
                       \int \frac{d^{3} k}{(2\pi)^{3}} \,
         f(-i\omega_n,{\bf k}) \;,
\ee
where $T$ is the temperature and
$\omega_n= 2 \pi n T$, $n= 0, \pm 1, \pm 2, \ldots$ 
are the bosonic Matsubara frequencies.
Our units are $\hbar=c=k_{B}=1$.  The metric tensor is $g^{\mu \nu}
= {\rm diag}(+,-,-,-)$.

\section{The $O(N)$ model in the CJT formalism}\label{sectionII}

In this section we discuss the application of the CJT formalism
to the $O(N)$ model. In Sec.\ \ref{results}
we shall present the numerical results for the
case $N=4$. The Lagrangian of the $O(N)$ linear sigma
model is given by 
\be
{\cal L}({\mbox{\boldmath$\phi$}})=
\frac 12\partial_\mu{\mbox{\boldmath$\phi$}}\cdot\partial^\mu
\mbox{\boldmath$\phi$}-U(\mbox{\boldmath$\phi$})
\ee
where $\mbox{\boldmath$\phi$}\equiv(\phi_1,\mbox{\boldmath$\pi$})$, 
with the first component $\phi_1$ corresponding to the scalar 
$\sigma$-meson and the other components 
$\mbox{\boldmath$\pi$}=(\phi_2,...,\phi_N)$
corresponding to the pseudoscalar pions. 
The function $U(\mbox{\boldmath$\phi$})$ is the tree-level potential,
\be \label{U}
U(\mbox{\boldmath$\phi$})=\frac 12\mu^2\mbox{\boldmath$\phi$}\cdot
\mbox{\boldmath$\phi$}
+\frac \lambda N(\mbox{\boldmath$\phi$}\cdot\mbox{\boldmath$\phi$})^2
-H\phi_1\,\,.
\ee
The constant $\mu^2$ is the bare mass and
$\lambda>0$ is the four-point coupling constant.
For $\mu^2 <0$ the $O(N)$ symmetry is spontaneously broken to $O(N-1)$,
leading to $N-1$ Goldstone bosons, the pions. The parameter $H$ breaks 
the symmetry explicitly, giving a mass to the pion.
The determination of the parameters $\mu^2, \lambda,$ and $H$ as
functions of the vacuum masses of $\sigma$-meson and pion, as well as
the pion decay constant, is discussed in detail in Appendix \ref{AppII}.

We assume translational invariance so that we may consider
the effective potential $V$ instead of the effective action
(\ref{effact}). For translationally invariant systems, these two
quantities are related via
\begin{equation}
\Gamma[\bar{\sigma},\bar{\mbox{\boldmath$\pi$}}, \bar{S},\bar{P}] 
= - \frac{\Omega_3}{T} \, 
V [\bar{\sigma},\bar{\mbox{\boldmath$\pi$}}, \bar{S},\bar{P}] \;,
\end{equation}
where $\Omega_3$ is the 3-volume of the system, and 
$\bar{\sigma}, \bar{\mbox{\boldmath$\pi$}}, \bar{S}, \bar{P}$ 
are the expectation values of the one- and two-point functions for 
the scalar and pseudoscalar fields in the presence of external sources
\cite{Cornwall:1974vz}.
We are interested in the case where these sources are zero, i.e.,
in the stationary points of $\Gamma$ or $V$, cf.\ Eq.\ (\ref{stat}).
Because the vacuum of QCD has even parity, the expectation values of the 
pseudoscalar fields are zero, $\bar{\mbox{\boldmath$\pi$}} = 0$,
and we shall simply omit the dependence of $V$ on $\bar{\mbox{\boldmath$\pi$}}$
in the following.

\begin{figure}
\includegraphics[height=5cm]{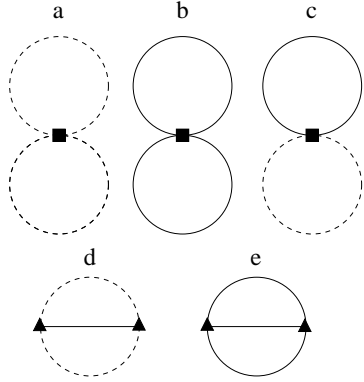}
\caption{The set of two-particle irreducible diagrams 
defining the improved Hartree-Fock approximation for the $O(N)$ model, see 
Eq.~(\ref{CJT-potential}). The diagrams a,b, and c are the double-bubble
diagrams, and d and e are the sunset diagrams. A full line denotes the
full propagator for the $\sigma$-meson and a dashed line the full
propagator for the pion. The four-particle vertex 
$\sim \lambda$ is represented by a square and the three-particle
vertex $\sim \lambda\sigma$  by a triangle.}
\label{paper1}
\end{figure}

Then, the effective potential for the $O(N)$ model in the CJT formalism reads
\cite{Roder:2003uz,Lenaghan:1999si}
\bea\label{CJT-potential}
V[\bar{\sigma},\bar{S},\bar{P}]
 & = & U(\bar{\sigma})  +  \half \int_Q \left[\, \ln \bar{S}^{-1}(Q) +
        S^{-1}(Q;\bar{\sigma})\, \bar{S}(Q)-1 \, \right] \nn \\
 & + &  \frac{N-1}{2} \int_Q \, \left[\,
        \ln \bar{P}^{-1}(Q) + P^{-1}(Q;\bar{\sigma})\, \bar{P}(Q)-
        1 \, \right]  +  V_{2}[\bar{\sigma},\bar{S},\bar{P}]\,\, ,
\eea 
where $ U(\bar{\sigma})$
is the tree-level potential (\ref{U}), evaluated at
$\mbox{\boldmath$\phi$}=(\bar{\sigma},0,\ldots,0)$.
The quantities $S^{-1}$ and $P^{-1}$ are the inverse tree-level
propagators for scalar and pseudoscalar mesons,
\bse
\bea  \label{Dsigma} 
S^{-1}(K;\bar{\sigma}) & = & -K^{2} + m_\sigma^2(\bar{\sigma})\; , \\
P^{-1}(K;\bar{\sigma}) & = & -K^{2} + m_\pi^2 (\bar{\sigma})\; , 
\label{Dpi}
\eea
\ese
where the tree-level masses are
\bse
\bea\label{tree_level_masses}
m_\sigma^2(\bar{\sigma})  & = & \mu^{2} + 
        \frac{12\, \lambda}{N}\, \bar{\sigma}^{2}\; , \\
m_\pi^2(\bar{\sigma}) &  =  &   \mu^2 +
        \frac{4\, \lambda}{N}\, \bar{\sigma}^{2}\; . 
\eea
\ese
The functional $V_2$ in Eq.\ (\ref{CJT-potential}) is related to 
$\Gamma_2$ in Eq.\ (\ref{effact}) via
$\Gamma_2 = - (\Omega_3/T)\, V_2$.
The Hartree-Fock approximation is defined by restricting the
sum of all 2PI diagrams, $V_2$, to 
the three double-bubble diagrams shown in Figs.~\ref{paper1} a, b, and c,
\bse
\bea \label{V2a}
V_{2}^a[\bar{P}] &\equiv&
        (N+1)(N-1)\, \frac{\lambda}{N} \left[
        \int_Q \, \bar{P}(Q)\right]^{2}\,\,,\\
V_{2}^b[\bar{S}] &\equiv& 3\, \frac{\lambda}{N}
        \left[ \int_Q \, \bar{S}(Q) \right]^{2}\,\,, \\
V_{2}^c[\bar{S},\bar{P}]&\equiv& 2\,(N-1) \frac{\lambda}{N}
        \int_Q\, \bar{S}(Q) 
        \int_L\, \bar{P}(L) \,\,.
\eea
\ese 
As explained in the introduction,
in order to include the nonzero decay width of the particles
we have to go beyond the Hartree-Fock approximation by
additionally including the sunset diagrams of
Figs.~\ref{paper1} d and e.
These diagrams have an explicit dependence on $\bar{\sigma}$, 
\bse
\bea 
V_2^d[\bar\sigma,\bar{S},\bar{P}]&\equiv&\frac{1}{2} \, 2(N-1)
\left(\frac{4\lambda \bar\sigma}{N}\right)^2\int_L\int_Q
\bar{S}(L)\bar{P}(Q)\bar{P}(L+Q)\; ,\\
V_2^e[\bar\sigma,\bar{S}]&\equiv&\frac{1}{2}\, 3!
\left(\frac{4\lambda \bar\sigma}{N}\right)^2\int_L\int_Q
\bar{S}(L)\bar{S}(Q)\bar{S}(L+Q)\; . \label{V2e}
\eea 
\ese
The complete expression for $V_2$ is obtained by the sum of
the contributions (\ref{V2a}) -- (\ref{V2e}),
\be
V_2=V_{2}^a+V_{2}^b+V_{2}^c+V_2^d+V_2^e\;.
\ee
The expectation values of the one- and two-point functions in 
the absence of external sources, $\sigma$ and
${\cal S}$, ${\cal P}$,
are determined from the stationary points of $V$,
\bea
\left.\frac{\delta V}{\delta \bar\sigma}\right
|_{\bar\sigma=\sigma,\bar{S}={\cal S},\bar{P}={\cal P}}=0\;,\;\;\;
\left.\frac{\delta V}{\delta \bar{S}}\right
|_{\bar\sigma=\sigma,\bar{S}={\cal S},\bar{P}={\cal P}}=0\;,\;\;\;
\left.\frac{\delta V}{\delta \bar{P}}\right
|_{\bar\sigma=\sigma,\bar{S}={\cal S},\bar{P}={\cal P}}=0\;,
\eea
leading to an equation for the scalar condensate $\sigma$,
\bse\label{gap_equations}
\be\label{condensate} 
H  =  \mu^2 \, \sigma + \frac{4 \lambda}{N}\, \sigma^3
 +  \frac{4\lambda}{N}\, \sigma \int_Q\, 
\left[ 3 \,{{\cal S}}(Q) + (N-1) \, {{\cal P}}(Q)\right] 
+ \left. \frac{\delta V_2}{\delta \bar\sigma}\right
|_{\bar\sigma=\sigma,\bar{S}={\cal S},\bar{P}={\cal P}}\; ,
\ee
where
\bea
\left. \frac{\delta V_2}{\delta \bar\sigma}\right
|_{\bar\sigma=\sigma,\bar{S}={\cal S},\bar{P}={\cal P}} & =&
\left(\frac{4\lambda}{N}\right)^2\sigma
\left[2(N-1)
\int_L\int_Q
{\cal S}(L){\cal P}(Q){\cal P}(L+Q) \right. \nn \\
&  & \hspace*{1.8cm}+ \left. 3!\int_L\int_Q
{\cal S}(L){\cal S}(Q){\cal S}(L+Q)\right]\; , \label{dv2ds}
\eea
and to the Dyson-Schwinger equations for the scalar and pseudoscalar
propagators,
\bea 
{\cal S}^{-1}(K;\sigma)&=&S^{-1}(K;\sigma)
+\Sigma[K;\sigma]\label{schwinger_dyson_1}\;,\\
{\cal P}^{-1}(K;\sigma)&=&P^{-1}(K;\sigma)
+\Pi[K;\sigma]\label{schwinger_dyson_2}\;.
\eea
\ese
Here we introduced the self-energies 
of the scalar and pseudoscalar fields,
\bse
\bea
\Sigma & = & \Sigma^a+\Sigma^b+\Sigma^c+\Sigma^d+\Sigma^e\;, \\
\Pi&=&\Pi^a+\Pi^b+\Pi^c+\Pi^d+\Pi^e\;,
\eea
\ese
with
\bse\label{self_energy_sigma}
\bea
\label{self_sigmaa}
\Sigma^a&\equiv&2\left.\frac{\delta V_2^a}{\delta \bar{S}}\right
|_{\bar\sigma=\sigma,\bar{S}={\cal S},\bar{P}={\cal P}}
=0\; ,\\
\label{self_sigmab}
\Sigma^b&\equiv&2\left.\frac{\delta V_2^b}{\delta \bar{S}}\right
|_{\bar\sigma=\sigma,\bar{S}={\cal S},\bar{P}={\cal P}}
=\frac{4\lambda}{N} \,  3\, \int_Q \, {\cal S}(Q)\; ,\\
\label{self_sigmac}
\Sigma^c&\equiv&2\left.\frac{\delta V_2^c}{\delta \bar{S}}\right
|_{\bar\sigma=\sigma,\bar{S}={\cal S},\bar{P}={\cal P}}
=\frac{4\lambda}{N} (N-1)\, \int_Q \, {\cal P}(Q)\; ,\\
\label{self_sigmad}
\Sigma^d[K;\sigma]&\equiv&2\left.\frac{\delta V_2^d}{\delta \bar{S}}\right
|_{\bar\sigma=\sigma,\bar{S}={\cal S},\bar{P}={\cal P}}
=\left(\frac{4\lambda \sigma}{N}\right)^2
2(N-1)\int_Q{\cal P}(K-Q){\cal P}(Q)\; ,\\
\label{self_sigmae}
\Sigma^e[K;\sigma]&\equiv&2\left.\frac{\delta V_2^e}{\delta \bar{S}}\right
|_{\bar\sigma=\sigma,\bar{S}={\cal S},\bar{P}={\cal P}}
=\left(\frac{4\lambda \sigma}{N}\right)^2
3\cdot 3!\int_Q{\cal S}(K-Q){\cal S}(Q)\; ,
\eea\ese
and
\bse\label{self_energy_pion}
\bea
\label{self_pia}
\Pi^a&\equiv&\frac{2}{N-1}\left.\frac{\delta V_2^a}{\delta \bar{P}}\right
|_{\bar\sigma=\sigma,\bar{S}={\cal S},\bar{P}={\cal P}}
=\frac{4\lambda}{N}\, (N+1)\,  \int_Q \, {\cal P}(Q)\; ,\\
\label{self_pib}
\Pi^b&\equiv&\frac{2}{N-1}\left.\frac{\delta V_2^b}{\delta \bar{P}}\right
|_{\bar\sigma=\sigma,\bar{S}={\cal S},\bar{P}={\cal P}}
=0\; ,\\
\label{self_pic}
\Pi^c&\equiv&\frac{2}{N-1}\left.\frac{\delta V_2^c}{\delta \bar{P}}\right
|_{\bar\sigma=\sigma,\bar{S}={\cal S},\bar{P}={\cal P}}
=\frac{4\lambda}{N} \int_Q \, {\cal S}(Q)\; ,\\
\label{self_pid}
\Pi^d[K;\sigma]&\equiv&\frac{2}{N-1}\left.\frac{\delta V_2^d}{\delta 
\bar{P}}\right
|_{\bar\sigma=\sigma,\bar{S}={\cal S},\bar{P}={\cal P}}
=\left(\frac{4\lambda \sigma}{N}\right)^2
4\int_Q{\cal P}(K-Q){\cal S}(Q)\;,\\
\label{self_pie}
\Pi^e&\equiv&\frac{2}{N-1}\left.\frac{\delta V_2^e}{\delta \bar{P}}\right
|_{\bar\sigma=\sigma,\bar{S}={\cal S},\bar{P}={\cal P}}=0\; .
\eea
\ese
The calculation of the self-energy contributions  
Eqs.~(\ref{self_energy_sigma}) and (\ref{self_energy_pion}) corresponds to
opening internal lines in the diagrams of Fig.~\ref{paper1}. Employing this
procedure to the 
double-bubble diagrams leads to the tadpole contributions
for the self-energies of the
$\sigma$-meson (see Figs.~\ref{selfenergy_sigma} b, c)
and the pion (see Figs.~\ref{selfenergy_pion} a, c).
This defines the Hartree-Fock approximation.
Additionally, the sunset diagrams of 
Figs.~\ref{paper1} d, e lead to the one-loop contributions shown
in Figs.~\ref{selfenergy_sigma} d, e and \ref{selfenergy_pion} d. 
The latter contributions depend on the energy and the momentum 
of the particles and lead to nonvanishing imaginary parts for the 
self-energies.
The explicit calculation of the self-energies (\ref{self_energy_sigma})
and (\ref{self_energy_pion})
is discussed in Appendix \ref{AppendixI}.

\section{Improving Hartree-Fock by including nonzero decay widths}\label{III}

On the basis of the general framework presented in the 
preceding section we now discuss in detail how the Hartree-Fock approximation
can be improved by including the nonzero decay width of the particles.

\begin{figure}
\includegraphics[height=2cm]{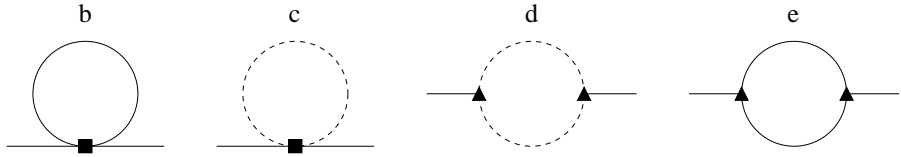}
\caption{The self-energy of the $\sigma$-meson. The diagrams b and c
are tadpole contributions generated by cutting an internal $\sigma$-line in the
double-bubble diagrams b and c in Fig.\ \ref{paper1}. The diagrams 
d and e are one-loop contributions generated by cutting an internal
$\sigma$-meson line in the sunset diagrams d and e of Fig.\ \ref{paper1}. 
Lines and vertices as in Fig.\ \ref{paper1}.}
\label{selfenergy_sigma}
\end{figure}

As explained in the introduction, we rewrite the set of Dyson-Schwinger
equations (\ref{schwinger_dyson_1}) and (\ref{schwinger_dyson_2}) in
terms of a set of integral equations for the spectral densities.
In general, the spectral densities for the $\sigma$-meson and the pion
are defined as
\bse\bea
\rho_\sigma (\omega, {\bf k}) &\equiv&2\, 
\Im\;{\cal S}(\omega+i\eta,{\bf k};\sigma)\;,\\
\rho_\pi (\omega, {\bf k}) &\equiv&2\, 
\Im\; {\cal P}(\omega+i\eta,{\bf k};\sigma)\;.
\eea\ese
If the imaginary parts of the self-energies are zero, as in
the Hartree-Fock approximation, all particles are stable quasiparticles,
i.e., their spectral densities become delta-functions with support
on the quasiparticle mass shell,
\bse \label{specdens1} \bea
\rho_\sigma (\omega, {\bf k}) &=& 2 \pi\, 
Z_\sigma[\omega_\sigma({\bf k}),{\bf k}]\, 
 \left\{ \delta [ \omega - \omega_\sigma({\bf k})] - \delta [ \omega + 
\omega_\sigma({\bf k})] \right\}\;,\\
\rho_\pi (\omega, {\bf k}) &=& 2\pi \, 
Z_\pi[\omega_\pi({\bf k}),{\bf k}]\, 
 \left\{ \delta [ \omega - \omega_\pi({\bf k})] - \delta [ \omega + 
\omega_\pi({\bf k})] \right\}\;,
\eea \ese
where $\omega_{\sigma,\pi}({\bf k})$ are the quasiparticle energies
for $\sigma$-mesons and pions, defined by the positive solutions of
\bse \label{qpe} \bea
\omega_\sigma^2({\bf k}) & = & k^2 + m_\sigma^2(\sigma) + 
\Re\, \Sigma[\omega_\sigma({\bf k}),{\bf k};\sigma]\;,\\
\omega_\pi^2({\bf k}) & = & k^2 + m_\pi^2(\sigma) + 
\Re\, \Pi[\omega_\pi({\bf k}),{\bf k};\sigma]\;,
\eea \ese
and 
\bse \bea
Z_{\sigma}[\omega_\sigma({\bf k}),{\bf k}] & \equiv & \left| 
\frac{\partial {\cal S}^{-1}(K;\sigma)}{\partial k_0} 
\right|_{k_0 = \omega_\sigma({\bf k})}^{-1}\;,\\
Z_{\pi}[\omega_\pi({\bf k}),{\bf k}] & \equiv & \left| 
\frac{\partial {\cal P}^{-1}(K;\sigma)}{\partial k_0} 
\right|_{k_0 = \omega_\pi({\bf k})}^{-1}\;,
\eea \ese
are the wave-function renormalization factors on the quasi-particle
mass shell. Since the real parts of the self-energies are even
functions of energy, these factors are also even in 
$\omega_{\sigma,\pi}({\bf k})$.
In the Hartree-Fock approximation, the (real parts of the) self-energies
do not depend on $K^\mu$, such that
\be \label{qpe2}
\omega_\sigma(k) = \sqrt{k^2 + M_\sigma^2(\sigma)}\;,
\;\;\;\;
\omega_\pi(k) = \sqrt{k^2 + M_\pi^2(\sigma)}\;,
\ee
where 
\be \label{effmass}
M_\sigma^2 (\sigma) \equiv m_\sigma^2(\sigma) + \Re\, \Sigma(\sigma)\;, 
\;\;\;\;
M_\pi^2 (\sigma) \equiv m_\pi^2(\sigma) + \Re\, \Pi(\sigma)\;,
\ee
are the effective $\sigma$-meson and pion masses, and
\be 
Z_{\sigma}[\omega_\sigma(k)] = \frac{1}{2 \omega_\sigma(k)}\;,
\;\;\;\;
Z_{\pi}[\omega_\pi(k)] = \frac{1}{2 \omega_\pi (k)}\;.
\ee
For nonvanishing imaginary parts of the self-energies the spectral
densities assume the following form:
\bse\label{spectral_density}
\bea
\rho_\sigma(\omega, {\bf k})&=&-\frac{2\,\Im\,\Sigma(\omega, {\bf k};\sigma)}
{[\omega^2-k^2-m_\sigma^2(\sigma)-\Re\, \Sigma(\omega, {\bf k};\sigma)]^2
+[\Im \,\Sigma(\omega, {\bf k};\sigma)]^2}\,\, , \\
\rho_\pi(\omega, {\bf k})&=&-\frac{2\,\Im\, \Pi(\omega, {\bf k};\sigma)}
{[\omega^2-k^2-m_\pi^2(\sigma)-\Re\, \Pi(\omega, {\bf k};\sigma)]^2
+[\Im\, \Pi(\omega, {\bf k};\sigma)]^2} \,\,.
\eea
\ese
The general calculation of the imaginary parts of the self-energies 
is discussed in the second part of Appendix \ref{AppendixI}. It turns out that
they do not depend on the direction of 3-momentum ${\bf k}$, but
only on the modulus $k$. The tadpole diagrams of 
Figs.~\ref{selfenergy_sigma} b, c (for the $\sigma$-meson) and
Figs.~\ref{selfenergy_pion} a, c (for the pion) do not have an
imaginary part. Thus [see Appendix \ref{AppendixI}, Eq.\ (\ref{general_im})],
\bse\label{im}\bea
\Im\,\Sigma(\omega, k ;\sigma)
&=&\Im \,\Sigma^d(\omega,k;\sigma)+\Im\, \Sigma^e(\omega,k;\sigma)\nn\\
&=&\left(\frac{4\lambda \sigma}{N}\right)^2
\frac{1}{2(2\pi)^3}\frac{1}{k}
\int_{-\infty}^{\infty} d \omega_1\, d \omega_2\,
[1+f(\omega_1)+f(\omega_2)]\delta(\omega-\omega_1-\omega_2)\nn \\
&\times& 
\int_0^{\infty}   d q_1 \, q_1 \, d q_2 \, q_2\;
\Theta\left(|q_1-q_2|\leq k\leq q_1+q_2\right)
\nn\\
&\times& 
\left[2(N-1)\, \rho_\pi(\omega_1,q_1)\rho_\pi(\omega_2,q_2)
+3\cdot 3!\,\rho_\sigma(\omega_1,q_1)\rho_\sigma(\omega_2,q_2)
\right]\,\,,\\
\Im\, \Pi(\omega,k;\sigma)&=&\Im\, \Pi^d(\omega,k;\sigma)\nn\\
&=&\left(\frac{4\lambda \sigma}{N}\right)^2
\frac{1}{2(2\pi)^3}\frac{1}{k}
\int_{-\infty}^{\infty} d \omega_1\, d \omega_2\,
[1+f(\omega_1)+f(\omega_2)]\delta(\omega-\omega_1-\omega_2)\nn \\
&\times& 
\int_0^{\infty}   d q_1 \, q_1 \, d q_2\, q_2\; 
\Theta\left(|q_1-q_2|\leq k\leq q_1+q_2\right)
\nn\\
&\times&
4\,\rho_\sigma(\omega_1,q_1)\rho_\pi(\omega_2,q_2)
\,\,,
\eea\ese
where $f(\omega)= [ \exp(\omega/T)-1 ]^{-1}$ is the Bose-Einstein
distribution function.
These expressions are not ultraviolet divergent and thus do not
need to be renormalized. 

\begin{figure}
\includegraphics[height=15cm]{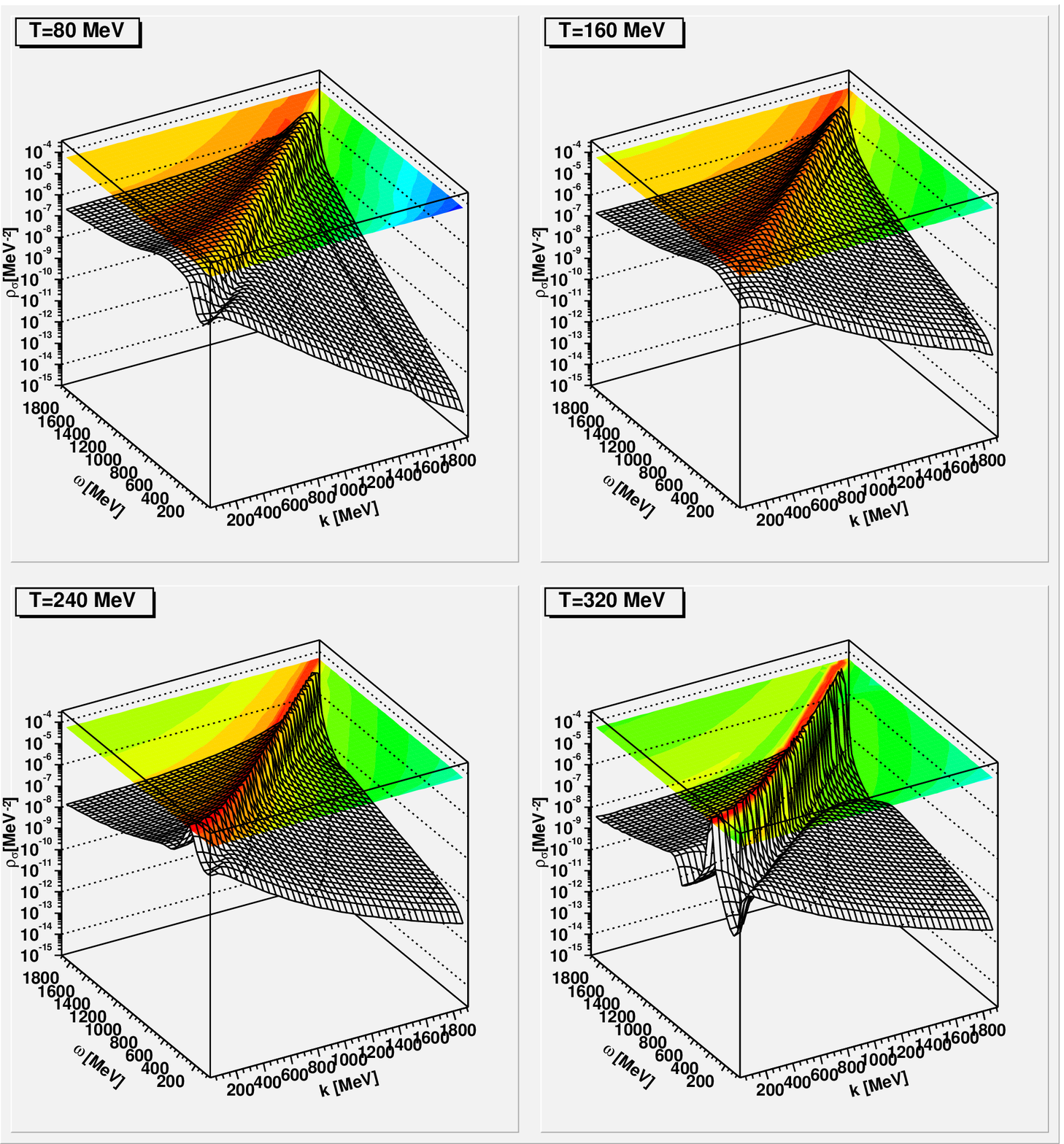}
\caption{The spectral density of the $\sigma$-meson
as a function of energy $\omega$ and momentum $k$
at temperatures 80, 160, 240, and 320 MeV.}
\label{rho_sigma}
\end{figure}

\begin{figure}
\includegraphics[height=15cm]{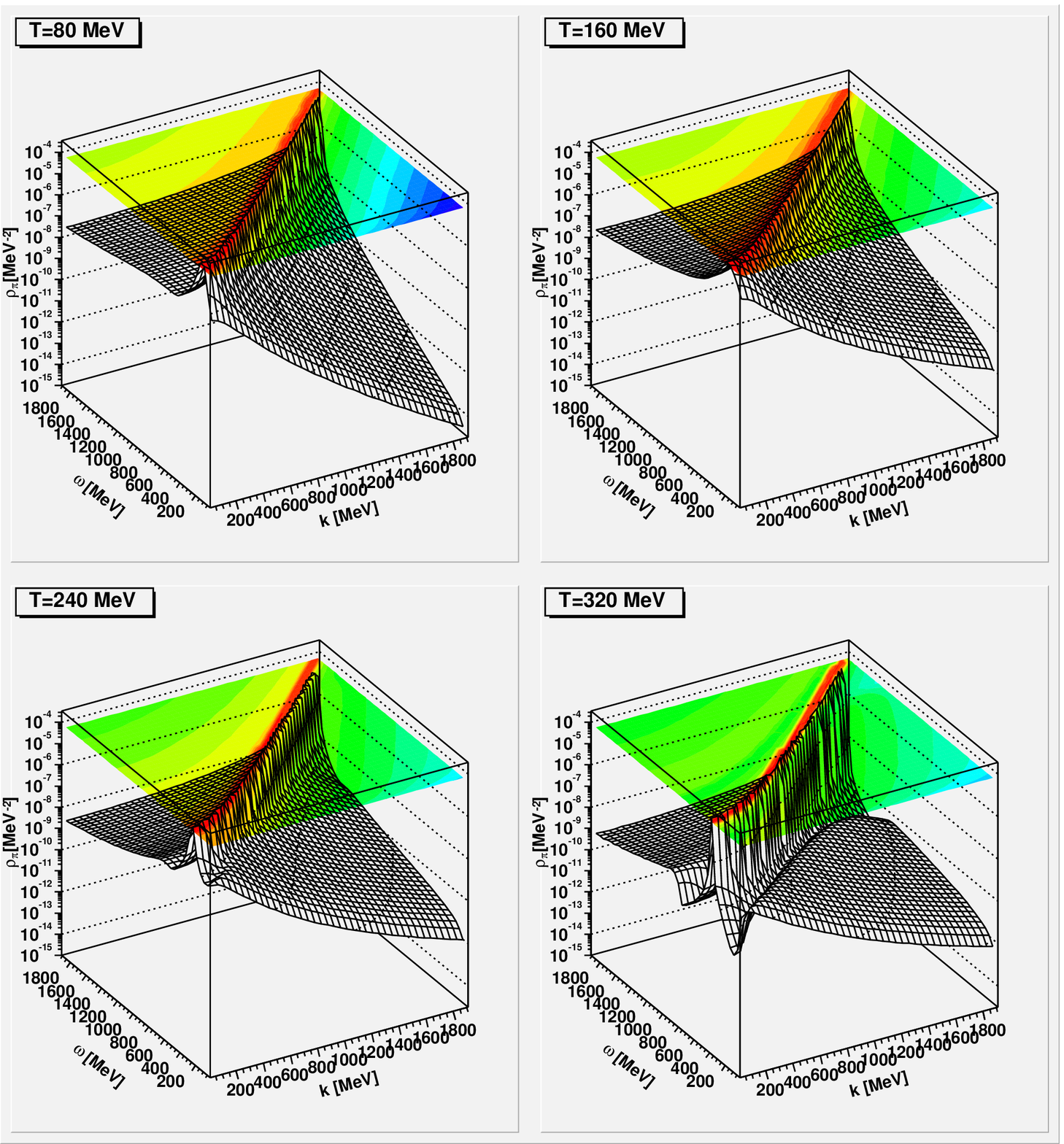}
\caption{The spectral density of the pion as a 
function of  energy $\omega$ and momentum $k$
at temperatures 
80, 160, 240, and 320 MeV.}
\label{rho_pion}
\end{figure}

For the real parts, we employ the following approximation. We only
take into account the Hartree-Fock contributions, arising 
from the tadpole diagrams Figs.~\ref{selfenergy_sigma} b, c (for the
$\sigma$-meson) and Figs.~\ref{selfenergy_pion} a, c (for the pion).
As discussed above, these contributions do not depend on energy and
momentum. The real parts of the diagrams in Figs.~\ref{selfenergy_sigma}
d, e and \ref{selfenergy_pion} d are functions of energy and momentum
and are much harder to compute, involving an additional numerical integration
as compared to the respective imaginary parts (\ref{im}).
This has been done in the Hartree (or large-$N$) approximation
to the $O(N)$ linear sigma model in Ref.\ \cite{Roder:2005qy}.
In this approximation, the only
energy-momentum dependent contribution arises
from the diagram in Fig.~\ref{selfenergy_sigma} d. 
It was found in Ref.\ \cite{Roder:2005qy}
that this contribution leads to minor quantitative, but not
qualitative, changes of the spectral density of the $\sigma$-meson.
Therefore, in the present first step where we focus exclusively on
the effects of a nonzero decay width, we neglect the
real parts of the diagrams in Figs.~\ref{selfenergy_sigma}
d, e and \ref{selfenergy_pion} d.
Consequently [see Appendix \ref{AppendixI}, Eq.~(\ref{general_eq_tadpole})],
\bse \label{re} \bea
\Re\, \Sigma(\sigma)&=&\Re\,\Sigma^b(\sigma)+\Re\, \Sigma^c(\sigma)\nn \\
&=&\frac{4\lambda}{N} \, 
\frac{2}{(2\pi)^3}
\int_{0}^{\infty} d \omega\, d q\, q^2\,[ 1 + 2\, f(\omega)]\,
\left[ 3\, \rho_\sigma(\omega,q) + (N-1) \, \rho_\pi(\omega,q)\right]\,,\\
\Re\, \Pi(\sigma)&=&\Re\, \Pi^a(\sigma)+\Re\, \Pi^c(\sigma)\nn \\
&=&\frac{4\lambda}{N} 
\frac{2}{(2\pi)^3}
\int_{0}^{\infty} d \omega\, d q\, q^2\,
[1 + 2\, f(\omega)]\, \left[ \rho_\sigma(\omega,q)
+ (N+1)\, \rho_\pi(\omega,q)\right]\,.
\eea\ese
Since the real parts do not depend on energy and momentum,
they only modify the masses of the $\sigma$-meson and the pion, 
cf.\ Eq.\ (\ref{effmass}). The thermal parts of the self-energies in
Eqs.\ (\ref{re}), $\sim f(\omega)$ are ultraviolet finite,
but the vacuum parts, $\sim 1$, are ultraviolet divergent and have to be 
renormalized. The $O(N)$ linear sigma model is an
effective theory which is not valid above a certain energy or
momentum scale. Therefore, we choose a particularly
simple renormalization scheme, regulating
the ultraviolet divergent part of the $\omega$ integral 
in Eqs.\ (\ref{re}) by introducing a cut-off $\Lambda$, cf.\ Appendix
\ref{AppendixI}.
In principle, we could also have chosen to regulate the $q$ integral. 
However, the spectral density
relates the energy to the momentum integral in Eqs.\ (\ref{re}).
For large momenta $q$, the spectral density is
of appreciable magnitude only in a range of energies $\omega \sim q$,
cf.\ Figs.\ \ref{rho_sigma} and \ref{rho_pion}.
Of course, our results
will depend on the cut-off $\Lambda$; this dependence
will be discussed in Sec.\ \ref{results}. 
The cut-off prescription is also used to compute the tadpole and the
sunset contributions in Eq.~(\ref{condensate}) for the condensate,
cf.\ Appendix \ref{AppendixI}.
For a more formal treatment of renormalization within the
CJT formalism, see for instance Refs.\
\cite{vanHees:2001ik,VanHees:2001pf,vanHees:2002bv}.

After obtaining real and imaginary parts (\ref{re}) and (\ref{im}), 
respectively, one inserts them in the expressions
(\ref{specdens1}) (if the imaginary part is zero) or
(\ref{spectral_density}) (if the imaginary part is nonzero) 
for the spectral densities. These spectral densities can then
be used to again evaluate the real and imaginary parts of the
self-energies. This defines an iterative scheme to self-consistently
solve for the spectral densities as functions of energy and momentum. 
A convenient starting point
for this scheme is the Hartree-Fock approximation, i.e., neglecting
the imaginary parts altogether.

The self-consistently computed spectral densities obey a sum rule
\cite{leBellac},
\be \label{sumrule}
\int_{-\infty}^\infty \frac{d \omega}{2 \pi} \, \omega\,
\rho_{\sigma,\pi}(\omega,{\bf k}) = 1\;.
\ee
In our calculation, there are two reasons why this
sum rule may be violated. One is because we neglected the
real parts of the self-energy diagrams in Figs.~\ref{selfenergy_sigma}
d, e and \ref{selfenergy_pion} d. (The results of
Ref.\ \cite{Roder:2005qy} suggest, though, 
that this has minor impact on the sum rule.)
The other is due to the numerical realization of the
above described iterative scheme. 
Numerically, one has to solve for the spectral density on a finite,
discretized grid in energy-momentum space. If the imaginary part
of the self-energy becomes very small, the spectral density converges
towards a delta function. The support of the delta function may 
be located between grid sites, which causes a loss of spectral strength,
which in turn violates the sum rule.

Our prescription to restore the validity of the sum rule is the
following. We first check whether
the imaginary part is smaller than the grid spacing (in our calculations,
10 MeV in both energy and momentum direction) at the
location of the quasiparticle mass shell, $\omega = \omega_{\sigma,\pi}(k)$.
If this is the case, we add a sufficiently wide numerical 
realization of the delta function, $\delta_{\rm num}$, 
to the original spectral density,
$\rho(\omega,k) \rightarrow \rho (\omega, k)
+ c \cdot \delta_{\rm num}[\omega - \omega_{\sigma,\pi}(k)]$,
where $c$ is a constant that is adjusted
so that the sum rule is fulfilled on our energy-momentum grid,
\be
\label{sumrule2}
\int_{-\omega_{\rm max}}^{\omega_{\rm max}} 
\frac{d \omega}{2 \pi} \, \omega\,
\rho_{\sigma,\pi}(\omega,{\bf k}) = 1\;,
\ee
where $\omega_{\rm max}$ is the maximum energy on the energy-momentum grid.

On the other hand, if the imaginary part turns out to be sufficiently
large, we presume that a possible violation
of the sum rule (\ref{sumrule2}) 
arises from neglecting the real parts of the
self-energy diagrams in Figs.~\ref{selfenergy_sigma}
d, e and \ref{selfenergy_pion} d. In this case, we
multiply the spectral density by a constant, $\rho \rightarrow c' \cdot
\rho$, where $c'$ is adjusted so that the sum rule (\ref{sumrule2}) is
fulfilled. (By comparing the results to the case $c'=1$, we 
found that this somewhat {\em ad hoc\/} correction procedure does not 
lead to major quantitative changes, in agreement with the results
of Ref.\ \cite{Roder:2005qy}.)

Restricting the sum rule to a finite range in energy as
in Eq.\ (\ref{sumrule2}), however, causes the following problem.
If the decay width of the particles is very large and
consequently the spectral density a rather broad distribution
around the quasiparticle mass shell, there will be parts 
which lie outside the energy-momentum grid. We could estimate the
magnitude of this physical effect if we knew the behavior of the 
spectral density at energies $\omega > \omega_{\rm max}$. 
This is possible at zero temperature (with the help of Weinberg's sum rules),
but not at nonzero temperature. Here, we simply
assume that this effect is sufficiently small to be neglected, i.e.,
we assume that a possible violation of the sum rule (\ref{sumrule2}) is
due to the two above mentioned artifacts and accordingly perform
the correction of the spectral density.

Another physical effect which causes a loss of spectral strength
is if the quasiparticle energy 
$\omega_{\sigma,\pi}(k) > \omega_{\rm max}$. 
In this case, we do not perform the correction of the 
spectral density as described above. This occurs at
large momenta $k$, close to the edge of the energy-momentum grid.
We checked that, 
in the numerical calculation of the integrals in Eqs.\ (\ref{im}),
the integrands become sufficiently small in this region, so that 
the imaginary parts are not sensitive to this effect.

This concludes the discussion of our improvement of the
Hartree-Fock approximation by self-consistently including 
nonzero decay widths.

\section{Results}\label{results}

In this section we present the numerical results for the $O(4)$
linear sigma model obtained in the 
Hartree-Fock approximation improved by including nonzero decay widths,
as discussed in Sec.\ \ref{III}.

The first step is to determine the parameters of the
model. This is discussed in detail in Appendix \ref{AppII}.
The zero-temperature values for the
mass of the pion and the scalar condensate are
$M_\pi=139.5$ MeV and $\sigma\equiv f_\pi=92.4$ MeV \cite{PDBook}.
The $\sigma$-meson has a large decay width because of its decay into two pions.
We take $M_\sigma = 600$ MeV and determine the width self-consistently
within our improved Hartree-Fock approximation.
For the cut-off required to renormalize the ultraviolet-divergent
contributions of loop integrals, we consider two choices,
$\Lambda = 0$ and $200$ MeV. (As discussed in Appendix \ref{AppII},
it is not possible to consider arbitrarily large
values for $\Lambda$; for $M_\sigma = 600$ MeV
the maximum value is $\Lambda_{\rm max}= 260$ MeV.)
For $\Lambda= 0$,
we obtain for the parameters of the $O(4)$ model the values
$H=(121.60\,\mbox{MeV})^3,\lambda=19.943$, and
$\mu^2=-(388.34\,\mbox{MeV})^2$, while for $\Lambda = 200$ MeV,
we have $H=(107.16\,\mbox{MeV})^3,\lambda=20.303$, and
$\mu^2=-(411.38\,\mbox{MeV})^2$.

\begin{figure}
\includegraphics[height=8cm]
{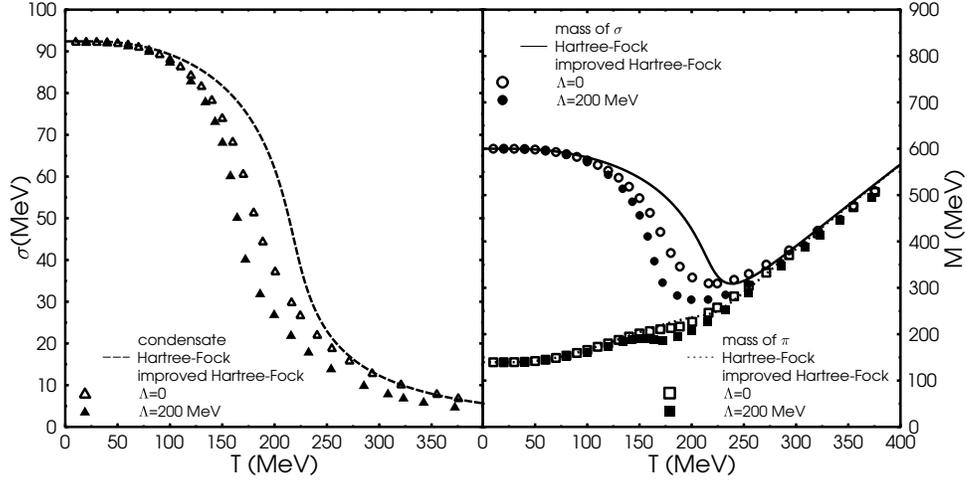}
\caption{The values for the condensate (left panel) 
and the effective masses of the $\sigma$-meson and the pion, (right panel)
as functions of temperature. The values are calculated in the Hartree-Fock
approximation (dashed and solid lines) 
and in the improved Hartree-Fock approximation (symbols)
as discussed in Sec.\ \ref{III}.}
\label{Hartree_values}
\end{figure}

The condensate $\sigma$ is shown in the left part of Fig.~\ref{Hartree_values}
as a function of temperature for the standard 
and the improved Hartree-Fock approximation. 
The qualitative behavior is similar in the two approximations:
the condensate drops significantly with temperature indicating the
restoration of chiral symmetry. Since we consider the case of explicit
$O(4)$ symmetry breaking by taking $H \neq 0$ in Eq.\ (\ref{U}),
the chiral phase transition is a cross-over transition.
Nevertheless, one can define a transition temperature, $T_\chi$,
as the temperature where the chiral susceptibility 
$\partial \sigma / \partial T$ assumes a maximum. 
Quantitatively, the inclusion of a nonzero
decay width lowers $T_\chi$ as compared to the standard 
Hartree-Fock approximation by about 
20\% for $\Lambda= 0$, and by about 25\% for $\Lambda = 200$ MeV, 
respectively. The transition temperature $T_\chi \simeq 160 - 175$~MeV
agrees well
with recent lattice results, $T_\chi \simeq 172 \pm 5$~MeV 
for the two-flavor case \cite{Laermann:2003cv}.
(Note, however, that the latter value is extracted from an
extrapolation to the chiral limit, while our result is for the
case of explicit symmetry breaking, i.e., for nonzero quark masses.)

For the solution of the condensate equation (\ref{condensate})
the relative magnitude of the
contribution from the sunset diagrams, Eq.\ (\ref{dv2ds}),
is negligibly small, of order $\sim 10^{-4}$, and thus can be safely neglected.
In turn, not having to compute the integrals pertaining to
the double loop, cf.\ Eq.\ (\ref{sunset}), 
considerably speeds up the computation.

\begin{figure}
\includegraphics[height=8cm]{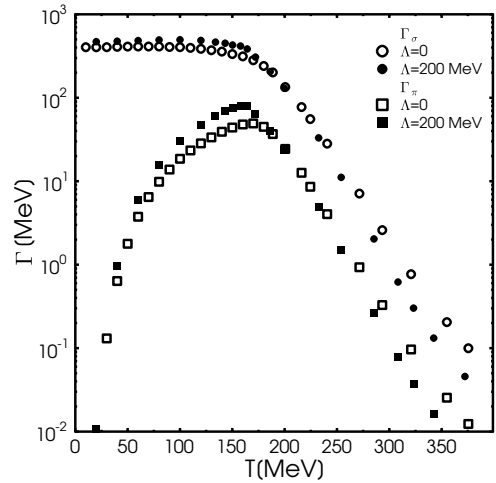}
\caption{The decay width of the $\sigma$-meson,
$\Gamma_\sigma=\Im \Sigma(\omega)/\omega$, and the pion, 
$\Gamma_\pi=\Im \Pi(\omega) /\omega$, in the
improved Hartree-Fock approximation as a function of the temperature 
at momentum $k=325~{\rm MeV}$.}
\label{decay_width}
\end{figure}

In the right part of Fig.~\ref{Hartree_values}, we show the effective masses 
$M_{\sigma}\equiv \sqrt{m_\sigma^2+\Re\, \Sigma}$
and $M_{\pi} \equiv \sqrt{m_\pi^2+\Re\, \Pi}$ of the $\sigma$-meson and 
the pion as functions of temperature in both approximation schemes. 
Since the decay width of the pion remains comparatively small, cf.\
Fig.~\ref{decay_width}, the mass of the pion does not change
appreciably when taking the nonzero decay width into account.
On the other hand, the large decay width of the $\sigma$-meson at
temperatures below $T_\chi$, cf.\ Fig.~\ref{decay_width}, 
does influence the mass: in this range of temperatures
the mass exhibits a stronger decrease with temperature in the
improved Hartree-Fock approximation.
At large temperatures $T > T_\chi$, the decay width of the $\sigma$-meson
becomes negligibly small, cf.\ Fig.~\ref{decay_width}, 
and the mass approaches the value computed in the standard
Hartree-Fock approximation.
Both $\sigma$-meson and pion masses approach each other
above $T_\chi$, indicating the restoration of chiral symmetry.

\begin{figure}
\includegraphics[height=15cm]{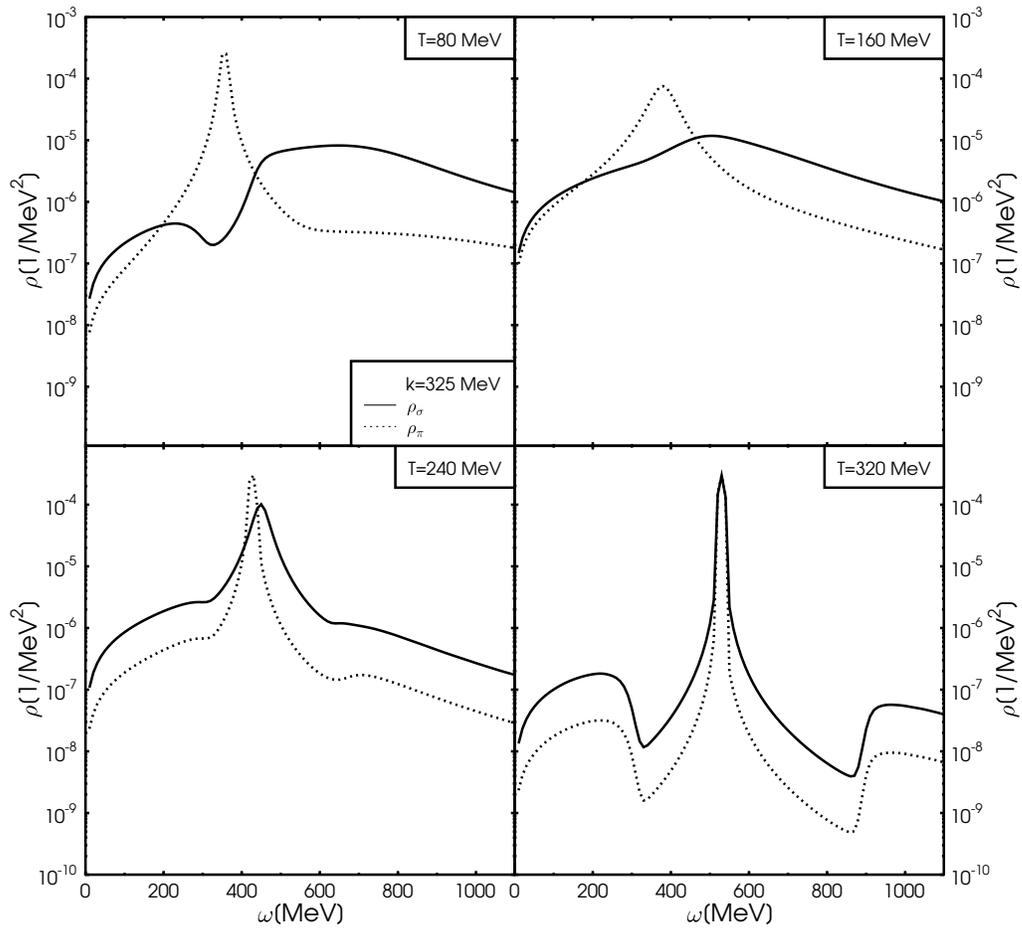}
\caption{The spectral density of the $\sigma$-meson and pion as a function
of energy $\omega$ at temperatures 80, 160,
240, and 320 MeV. The momentum is $k~=~325~{\rm MeV}$.}
\label{rho_sigma_pion}
\end{figure}

In Fig.~\ref{decay_width} we show the decay widths of $\sigma$-mesons
and pions, defined as \cite{Weldon:1983jn,leBellac}
\be
\Gamma_\sigma (k) \equiv
\frac{\Im\,\Sigma[\omega_\sigma(k),k;\sigma]}{
\omega_\sigma(k)}\;,\;\;\;\;
\Gamma_\pi (k) \equiv \frac{\Im\,\Pi[\omega_\pi(k),k;\sigma]}{
\omega_\pi(k)}\;,
\ee 
where $\omega_{\sigma,\pi}(k)$ is the energy on
the quasi-particle mass shell, cf.\ Eq.~(\ref{qpe}).
At small temperatures, due to the possible decay of a $\sigma$-meson
into two pions, the decay width of the former
is large, of the order of its mass.
Note that the values of $\Gamma_\sigma$ obtained here
at $T=0$ are completely determined by the parameters $m^2, \lambda,$ and $H$
of the $O(4)$ model, i.e., without adjusting any additional
parameter. They are reasonably close to the experimentally measured
value $\Gamma_\sigma \sim 600 - 1000$~MeV \cite{PDBook}.
The $\sigma$ decay width increases with temperature up to a maximum
at a temperature $T\simeq 100$ MeV and then
decreases rapidly. The decay width of the pion vanishes at
$T=0$. It increases at nonzero temperature
and assumes a maximum around $T\simeq 150$~MeV. At this temperature
the decay width is about half of the corresponding pion mass. It
decreases rapidly at higher temperature. Although the decay
widths of both particles decrease at high temperatures,
they do not become degenerate, the decay width of the $\sigma$-meson 
remains about a factor of 8 larger than that
of the pion. This difference can be traced to the symmetry factors
multiplying the one-loop self-energies for the $\sigma$-meson
and pion, cf.\ Eqs.\ (\ref{im}). For the $\sigma$-meson, there is a
factor $2 (N-1)$ in front of the pion loop and a factor 
$3 \cdot 3!$ in front of the $\sigma$-loop, so that the overall
factor is $\sim 2 (N-1) + 3 \cdot 3! = 24$. For the pion,
there is only a mixed $\sigma$-pion loop with a symmetry factor
$4$. From this simple argument
one already expects that the decay width of the $\sigma$-meson is about
a factor of 6 larger than that of the pion. The remaining difference
comes from the fact that the self-consistently computed
spectral densities of $\sigma$-meson and pion
under the integrals in Eqs.\ (\ref{im}) are also different, cf.\
Fig.\ \ref{rho_sigma_pion}. 

One might argue that, at asymptotically large temperatures,
effects from chiral symmetry breaking can no longer play a role
and the decay widths, as well as the spectral densities, 
of $\sigma$-meson and pion should become degenerate.
This is true in the chiral limit, where there is a 
thermodynamic phase transition between the phases of broken and
restored chiral symmetry, and where $\sigma \equiv 0$ in the latter phase.
Here, however, we consider the case of explicitly broken chiral symmetry,
where $\sigma >0 $, even when the temperature is very 
large. Since the decay widths are
proportional to $\sigma^2$, they also do not vanish at large temperature.
The difference in the symmetry factors for the self-energies
of $\sigma$-mesons and pions then leads to different
values for the decay widths and spectral densities.

The self-consistently calculated spectral densities of 
the $\sigma$-meson and the pion as functions 
of the external energy $\omega$ and 
momentum $k$ are shown in Fig.~\ref{rho_sigma} 
and Fig.~\ref{rho_pion} for different temperatures,
$T=80,\, 160,\, 240,$ and $320~\,{\rm MeV}$. Here we restrict
ourselves to the case $\Lambda = 0$; the results for
the case $\Lambda = 200$ MeV are quantitatively rather similar.
For a detailed discussion, let us fix the momentum
at $k = 325$ MeV and consider the spectral densities
as functions of energy $\omega$ for different temperatures,
cf.\ Fig.~\ref{rho_sigma_pion}.

At all temperatures, the pion spectral density (dotted line) 
exhibits a pronounced peak on
the mass shell, at $\omega_\pi(k) = \sqrt{k^2 + M_\pi^2(\sigma)}$.
When the temperature is below $200$ MeV, such that 
$M_\pi \simeq  139.5$ MeV, cf.\ Fig.~\ref{Hartree_values},
the peak is located at $\omega_\pi(325\, {\rm MeV}) \simeq 350$ MeV. 
Above $T \sim 200$ MeV, $M_\pi$ increases significantly with temperature, 
and the position of the peak shifts towards larger energies, 
$\omega_\pi(325\, {\rm MeV}) \simeq 500$ MeV.
The broadening of the peak is due to scattering of the pion
off $\sigma$-mesons in the medium.

In contrast to the pion spectral density, for temperatures
below $\sim 170$ MeV the $\sigma$
spectral density (full line) does not exhibit a peak at the
mass shell, $\omega_\sigma (k) = \sqrt{k^2 + M_\sigma^2(\sigma)}$.
The reason is that $\omega_\sigma(k)$ is still 
sufficiently large to allow for the decay into two pions.
Consequently, in this temperature range the $\sigma$ spectral
density is very broad. On the other hand,
for temperatures above $\sim 170$ MeV, where $\omega_\sigma(k)$
drops below $2 \, M_\pi$, the two-pion decay channel is
closed and the $\sigma$ spectral density develops a distinct peak,
whose width is due to scattering of the $\sigma$-mesons off pions and
other $\sigma$-mesons in the medium.

Two other features of the spectral densities shown in
Fig.~\ref{rho_sigma_pion} are noteworthy. The first is
the region below the light-cone, $K^2 =\omega^2- k^2 <0$, where 
the mesons are Landau-damped. The second is the 
two-particle decay threshold. For $\sigma \rightarrow \pi \pi$,
this threshold is located at $\omega \sim 2\, M_{\pi}$, for 
$\sigma \rightarrow \sigma \sigma$, it is at $\omega \sim 2\, M_{\sigma}$,
and for $\pi \rightarrow \sigma \pi$ at $\omega \sim M_\sigma + M_\pi$.
The threshold is most clearly seen at large temperatures, e.g.\ $T= 320$ MeV,
when both particles become degenerate in mass,
$M_\sigma \simeq M_\pi \sim 400$ MeV, and the threshold is at
$\omega \sim 900$ MeV.

\section{Conclusions}\label{conclusion}


In this paper we have studied the $O(N)$ linear sigma model 
at nonzero temperature within a self-consistent many-body resummation scheme.
This scheme extends the standard Hartree-Fock approximation by
including nonzero decay widths of the particles.
In the Hartree-Fock approximation, the self-energies of
the particles consist of tadpole diagrams which have no
imaginary part. Consequently, all particles are stable
quasi-particles.
In order to obtain a nonzero decay width, one has to include
diagrams in the self-energy, which have a nonzero imaginary part
corresponding to decay and, in a medium, scattering processes.

In order to incorporate the nonzero decay width in a self-consistent
way, we apply the Cornwall-Jackiw-Tomboulis formalism. 
The standard Hartree-Fock approximation is obtained by
considering only double-bubble diagrams in the 2PI effective action, leading
to the (energy- and momentum-independent)
tadpole contributions in the 1PI self-energies.
In order to extend the Hartree-Fock approximation, we additionally
take into account diagrams of sunset topology in the 
2PI effective action. This has the
consequence that the 1PI self-energies obtain additional energy- and
momentum-dependent one-loop contributions which have a nonzero imaginary part.
The spectral densities of $\sigma$-mesons and pions are then
computed as solutions of a self-consistent set of Dyson-Schwinger equations 
for the $\sigma$-meson and pion two-point functions,
coupled to a fix-point equation for the chiral condensate. We
only take into account the imaginary parts of the new one-loop
contributions. We made sure that the spectral densities obey a standard
sum rule by adjusting their normalization, if necessary.


We found that the temperature $T_\chi$ for chiral symmetry restoration 
is about $20 - 25 \%$ smaller as compared to the Hartree-Fock approximation 
when including nonzero particle decay widths. 
Our value for the temperature of
chiral symmetry restoration $T_\chi$ agrees reasonably well with 
lattice results \cite{Laermann:2003cv}. 
We computed the decay widths of $\sigma$-mesons and pions as
a function of temperature. The vacuum value for the $\sigma$ decay width
comes out to be in the experimentally observed range, without adjusting a 
parameter of the model. It stays approximately constant up
to temperatures $ \sim T_\chi$ and then decreases sharply with temperature.
The pion decay width grows from zero at $T=0$ to a value $\sim 100$ MeV at
$T \sim T_\chi$, and then also decreases with temperature.

We also investigated the spectral densities of $\sigma$-mesons and
pions as functions of energy and momentum for temperatures in the
range of $T=80$ to 320 MeV.
Below the chiral phase transition, the spectral density of the
$\sigma$-meson is broad, due to the possible decay into two pions.
It develops a peak at the quasi-particle mass shell above
the chiral phase transition, when this decay channel is closed.
On the other hand, the spectral density of the pion always exhibits a 
distinct peak at the quasi-particle mass shell. The width of this peak is 
due to scattering off $\sigma$-mesons in the medium.
Above the chiral phase transition, the spectral densities 
of $\sigma$-mesons and pions become degenerate in shape.


The present study can be continued along several lines.
An important task is to include also the real parts of
the self-energy contributions from the sunset diagrams.
It would also be of interest to study the chiral limit, in order
to see how the sunset diagrams affect the order
of the chiral phase transition. In the Hartree-Fock
approximation, the chiral phase transition turns out to be of
first order, while universality arguments predict it to be of 
second order.

Other possible future projects are the inclusion of other degrees
of freedom, for instance strange scalar and pseudoscalar
mesons, baryons \cite{beckmannphd}, and vector mesons
\cite{vanHees:2000bp,Ruppert:2004yg,strueberdipl}. 
The latter is of particular importance, since
in-medium changes in the spectral properties of vector mesons 
are reflected in the dilepton spectrum \cite{Rapp:1999ej} which, in turn, is 
experimentally observable in heavy-ion collisions
at GSI-SIS, CERN-SPS and BNL-RHIC energies.

\section*{Acknowledgments}
J.R.\ acknowledges support by the Alexander von Humboldt foundation as a
Feodor-Lynen fellow. We thank the Center of Scientific
Computing of the University of Frankfurt for computational support.
\appendix
\section{The calculation of the diagrams}\label{AppendixI}
\begin{figure}
\includegraphics[height=3.5cm]{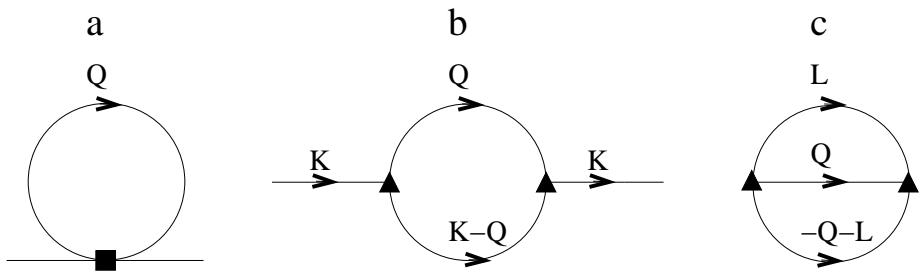}
\caption{The general topology of the tadpole diagram a, the
cut sunset diagram b, and the sunset diagram c.}
\label{diagrams}
\end{figure}

In this Appendix we discuss the calculation of the three types of
diagrams (see Fig.~\ref{diagrams}) contributing to the self-energies and the 
condensate equation. To evaluate them explicitly, we use 
standard techniques of 
thermal field theory, see for example Refs.~\cite{das,kapusta,leBellac}.

\subsection{The tadpole diagram}
In the following we discuss the calculation of the tadpole diagram
in Fig.~\ref{diagrams} a as a functional of the spectral density. 
In the imaginary-time formalism it is given by
\bea
{\cal T}\equiv T\sum_n \int \frac{d^3q}{(2\pi)^3} \Delta(-i\omega_n,{\bf q})\;,
\eea
where $T$ is the temperature, and $\Delta$ is either the $\sigma$-meson
or pion propagator. 

The first step is to perform the sum over the Matsubara frequencies.
To this aim we introduce the mixed representation of the propagator
\be\label{mixed_representation}
\Delta(-i\omega_n,{\bf q})
=\int_0^{1/T} d\tau\exp(-i\omega_n\tau)\Delta(\tau,{\bf q}),
\ee
with
\bea\label{def_spectral_density}
\Delta(\tau,{\bf q})=\int_{-\infty}^{\infty}
\frac{d \omega}{2\pi}\, [\Theta(\tau)+f(\omega)]\,\rho(\omega,{\bf q})
\exp(-\omega\tau)\,\,,
\eea
where $f(\omega)=[\exp(\omega/T)-1]^{-1}$ is the Bose-Einstein distribution
function. With the identity
\be\label{delta_representation}
T\sum_n\exp(-i\omega_n\tau)=\sum_{m=-\infty}^{\infty}
\delta\left(\tau-\frac{m}{T}\right)\; ,
\ee
the Matsubara sum can be performed analytically
\bea
{\cal T}&=&T\sum_{n} 
\int \frac{d^3q}{(2\pi)^3}
\int_0^{1/T} d\tau \exp(-i\omega_n\tau)
\Delta(\tau,{\bf q}) \nn\\
&=&\int \frac{d^3q}{(2\pi)^3}
\int_0^{1/T} d\tau 
\Delta(\tau,{\bf q})\left[\delta({\tau})+ \delta\left(\tau - \frac{1}{T}
\right) \right] \nn \\
& = & \int \frac{d^3q}{(2\pi)^3} \frac{1}{2} \left[
\Delta(0,{\bf q}) + \Delta\left(\frac{1}{T},{\bf q} \right) \right]
\nn \\
& = & \int \frac{d^3q}{(2\pi)^3} \Delta(0,{\bf q})\,\,,
\eea
where use has been made of the KMS condition $\Delta(\tau,{\bf q})
\equiv \Delta(\tau- 1/T, {\bf q})$. With 
Eq.~(\ref{def_spectral_density}) we finally have
\be\label{rm_T_1}
{\cal T}=
\int_{-\infty}^{\infty} \frac{d \omega}{2 \pi}
\int \frac{d^3q}{(2\pi)^3}
\left[\frac 12+f(\omega)\right]\rho(\omega,{\bf q})\,\,.
\ee
Due to isotropy of space,
the spectral density of a scalar particle
cannot depend on the direction of ${\bf q}$, thus
the angular integration can be carried out:
\bea
{\cal T}=4\pi
\int_{-\infty}^{\infty} \frac{d \omega}{2 \pi}
\int_{0}^{\infty}\frac{d\,q}{(2\pi)^3}\,q^2
\left[\frac 12+f(\omega)\right]\rho(\omega,q)\,\,.
\eea
Using the fact that the spectral density for bosonic
degrees of freedom is an odd function of the energy, 
$\rho(\omega)=-\rho(-\omega)$,
\bea\label{general_eq_tadpole}
{\cal T}&=&4\pi
\int_{0}^{\infty} \frac{d \omega}{2 \pi}\,
\frac{d q}{(2\pi)^3}\,q^2
[f(\omega)-f(-\omega)]\rho(\omega,q)\nn\\
&=&4\pi
\int_{0}^{\infty} \frac{d \omega}{2 \pi}\,
\frac{d q}{(2\pi)^3}\,q^2\,
[1+2f(\omega)]\,\rho(\omega,q)\,\,.
\eea
This integral can be split into an ultraviolet-finite 
temperature-dependent part,
\be
{\cal T}^T=4\pi
\int_{0}^{\infty} \frac{d \omega}{2 \pi}\,
\frac{d q}{(2\pi)^3}\,q^2\,2f(\omega)\,\rho(\omega,q)\,\,,
\ee
and an ultraviolet-divergent vacuum part, which
is renormalized by introducing a cut-off $\Lambda$ for
the $\omega$ integral,
\be\label{tadpole_lambda}
{\cal T}^\Lambda=4\pi
\int_{0}^{\infty} \frac{d \omega}{2 \pi}\,
\frac{d q}{(2\pi)^3}\,q^2\,\Theta(\Lambda - \omega)\,\rho(\omega,q)\,\,,
\ee
where ${\cal T}\equiv {\cal T}^T+{\cal T}^\Lambda$.

\subsection{The cut sunset diagram}\label{amputated_sunset}
In this section we calculate the imaginary part of the
cut sunset diagram of Fig.~\ref{diagrams} b. 
In the imaginary-time formalism 
the cut sunset diagram is given by
\bea
{\cal C}(-i\omega_m,{\bf k})\equiv
\int_Q \Delta_1 (K-Q)\, \Delta_2(Q)
=T\sum_n \int \frac{d^3q}{(2\pi)^3}
\Delta_1(- i(\omega_m-\omega_n),\mathbf{k-q})
\Delta_2(-i\omega_n,{\bf q})\,\,,
\eea
where $\Delta_1,\Delta_2$ are the propagators of $\sigma$-mesons and/or
pions. The major difference
between the tadpole diagram, discussed in the last section, and this diagram
is that it explicitly depends on the external four-momentum, 
$K^\mu\equiv(-i\omega_m,{\bf k})$.

Analogously to the last section one
introduces the mixed representation [see Eq.~(\ref{mixed_representation})] 
of the propagators $\Delta_1$ and $\Delta_2$ 
\bea
{\cal C}(-i\omega_m,{\bf k})
=T\sum_n \int \frac{d^3q}{(2\pi)^3}
&&\int_0^{1/T} d \tau_1\exp[-i(\omega_m-\omega_n)\tau_1]
\Delta_1(\tau_1,\mathbf{k-q})\nn\\
\times&&\int_0^{1/T} d \tau_2\exp(-i\omega_n\tau_2)
\Delta_2(\tau_2,{\bf q})\,\,.
\eea
The summation over the Matsubara frequencies leads to 
a delta function 
[cf.\ Eq.~(\ref{delta_representation})]
\bea
{\cal C}(-i\omega_m,{\bf k})
=\int\frac{d^3q}{(2\pi)^3}
\int_0^{1/T} d \tau\exp(-i\omega_m\tau)
\Delta_1(\tau,\mathbf{k-q})\Delta_2(\tau,{\bf q})
\,\,.
\eea
Introduction of the spectral densities
[see Eq.~(\ref{def_spectral_density})] for both
propagators in the mixed representation 
leads to
\bea
{\cal C}(-i\omega_m,{\bf k})
=\int \frac{d^3q}{(2\pi)^3}
\int_0^{1/T} d \tau\exp(-i\omega_m\tau)
&&\int_{-\infty}^{\infty}
\frac{d \omega_1}{2\pi}[1+f(\omega_1)]\rho_1(\omega_1,\mathbf{k-q})
\exp(-\omega_1\tau)\nn\\
\times&&\int_{-\infty}^{\infty}
\frac{d \omega_2}{2\pi}[1+f(\omega_2)]\rho_2(\omega_2,{\bf q})
\exp(-\omega_2\tau)\;.
\eea
The integration over $\tau$ can be performed analytically. 
Employing the identity
\bea
\left[\exp\left(-\frac{\omega_1+\omega_2}{T}\right)-1\right]
\,[1+f(\omega_1)][1+f(\omega_2)]
=- 1 - f(\omega_1)- f(\omega_2)\;,
\eea
the expression can be rewritten as 
\be
{\cal C}(-i \omega_m,{\bf k})=
\int_{-\infty}^{\infty} \frac{d \omega_1}{2 \pi} \,\frac{d \omega_2}{2 \pi} 
\frac{1+f(\omega_1)+f(\omega_2)}
{i\omega_{\rm m}+\omega_1+\omega_2}
\int \frac{d^3 q }{(2 \pi)^3}\, \rho_1(\omega_1,\mathbf{k-q})
\rho_2(\omega_2,{\bf q})\;. 
\ee
Using the fact that the 
spectral densities do not depend on the direction of momentum and
\bea\label{abs_ident_2}
\int \frac{d^3 x}{(2\pi)^3}
f(\abs{\mathbf{y-x}})g(x)
&=&\frac{2\pi}{y}\frac{1}{(2\pi)^3}
\int_0^{\infty} d x_1 \,  x_1 \, d x_2\, x_2\,
f(x_1)g(x_2)\nn\\
&\times& \Theta\left(|x_1-x_2|
\leq y\leq x_1+x_2\right)\,\,,
\eea
one can perform one angular integration, yielding
\bea
{\cal C}(-i \omega_m,{\bf k})&=&
\int_{-\infty}^{\infty} \frac{d \omega_1}{2 \pi} \,
\frac{d \omega_2}{2 \pi} \,
 \frac{1+f(\omega_1)+f(\omega_2)}{i\omega_{\rm m}+\omega_1+\omega_2}
 \\
&\times&
\frac{2\pi}{k} \frac{1}{(2\pi)^3} 
\int_0^{\infty}  d q_1\, q_1 \, d q_2 \, q_2\,
\rho_1(\omega_1,q_1)\rho_2(\omega_2,q_2)\nn\\
&\times& \Theta\left(|q_1-q_2|
\leq k\leq q_1+q_2\right)\,\,.\nn
\eea
The imaginary part of the retarded cut sunset diagram
can be extracted using analytical continuation, 
$-i \omega_m \rightarrow \omega + i \epsilon$, and
the Dirac identity $\Im 1/(x+i\epsilon)=-\pi\delta(x)$,
\bea\label{general_im}
\Im \,{\cal C}(\omega,{\bf k})&=&
\frac{1}{2(2\pi)^3}\frac{1}{ k}
\int_{-\infty}^{\infty} d \omega_1\, d \omega_2
[1+f(\omega_1)+f(\omega_2)]\, \delta(\omega-\omega_1-\omega_2)\nn \\
&\times& 
\int_0^{\infty}  d q_1 \, q_1 \, d q_2 \, q_2\, 
\Theta\left(|q_1-q_2|\leq k\leq q_1+q_2\right)\,
\rho_1(\omega_1,q_1)\, \rho_2(\omega_2,q_2)\,\,.
\eea
This integral is finite and therefore does not require renormalization.

\subsection{The sunset diagram}

In the imaginary-time formalism the sunset diagram shown in 
Fig.~\ref{diagrams} c is given by the following expression:
\bea
{\cal S}&=&T^2\sum_{n,m} 
\int \frac{d^3l}{(2\pi)^3}\, \frac{d^3q}{(2\pi)^3}
\Delta_1(-i\omega_m,{\bf l})
\Delta_2(-i\omega_n,{\bf q})
\Delta_3(-i(-\omega_n-\omega_m)
,-\mathbf{q-l})\;.
\eea
We introduce the mixed
representation, see Eq.~(\ref{mixed_representation}), for the three
propagators $\Delta_1,\Delta_2$, and $\Delta_3$. One can perform
the Matsubara sums employing Eq.~(\ref{delta_representation}). 
In the mixed representation one introduces the spectral densities 
$\rho_1,\rho_2$, and $\rho_3$
in order to perform the $\tau$ integration analytically.
Employing the identity 
\bea
&&\left[\exp\left(-\frac{\omega_1+\omega_2+\omega_3}{T} \right) -1\right]
\left[1+f(\omega_1)\right]
\left[1+f(\omega_2)\right]\left[1+f(\omega_3)\right]\nn\\
&=&- 1-f(\omega_1)-f(\omega_2)-f(\omega_3)
-f(\omega_1)f(\omega_2)-f(\omega_1)f(\omega_3)-f(\omega_2)f(\omega_3)\,\,,
\eea
for the Bose-Einstein distribution functions we obtain 
\bea 
{\cal S}&=&
\int_{-\infty}^{\infty} \frac{d \omega_1}{2 \pi}\,
\frac{d \omega_2}{2 \pi} \,
\frac{d \omega_3}{2 \pi}
\, \frac{1+f(\omega_1)+f(\omega_2)+f(\omega_3) 
+f(\omega_1)f(\omega_2)
+f(\omega_1)f(\omega_3)
+f(\omega_2)f(\omega_3)}
{\omega_1+\omega_2+\omega_3}  \nn \\
&\times& 
\int \frac{d^3 l }{(2 \pi)^3} \, \frac{\d^3 q }{(2 \pi)^3}
\, \rho_1(\omega_1,{\bf l}) \rho_2(\omega_2,{\bf q})
\rho_3(\omega_3,\mathbf{-q-l})\,\,.
\eea
Using the fact that the spectral densities do not depend on the
direction of momentum, the angular integrals
can be carried out with the result
\bea 
{\cal S}&=&\frac{2}{(2\pi)^7}
\int_{-\infty}^{\infty} d \omega_1\, d \omega_2\, d \omega_3
\int_0^{\infty} dq_1\, q_1\, dq_2\, q_2\, dq_3\, q_3\,
\Theta\left(|q_1-q_2|\leq q_3\leq q_1+q_2\right)\nn\\
&\times&
\frac{1+f(\omega_1)+f(\omega_2)+f(\omega_3) 
+f(\omega_1)f(\omega_2)+f(\omega_1)f(\omega_3)+f(\omega_2)f(\omega_3)
}{\omega_1+\omega_2+\omega_3}  \nn \\
&\times& 
\rho_1(\omega_1,q_1)
\rho_2(\omega_2,q_2)
\rho_3(\omega_3,q_3)
\,\,.
\eea
Using the antisymmetry of the spectral densities and relabelling
integration variables, we finally arrive at
\bea 
{\cal S}&=&\frac{4}{(2\pi)^7}
\int_{0}^{\infty} d \omega_1\,d \omega_2\,d \omega_3
\, dq_1\, q_1\, dq_2\, q_2\,dq_3\, q_3\,
\Theta\left(|q_1-q_2|\leq q_3\leq q_1+q_2\right)\nn\\
&\times&\left[\frac{f(\omega_1)f(\omega_2)+f(\omega_1)f(\omega_3)
          +f(\omega_3)f(\omega_2)+f(\omega_1)+f(\omega_2)
          +f(\omega_3)+1}{\omega_1+\omega_2+\omega_3}\right.\nn\\
&&\quad+\frac{f(\omega_1)f(\omega_3)
             -f(\omega_1)f(\omega_2)
             +f(\omega_2)f(\omega_3)
             +f(\omega_3)}{\omega_1+\omega_2-\omega_3}\nn\\
&&\quad+\frac{f(\omega_1)f(\omega_2)
             -f(\omega_1)f(\omega_3)
             +f(\omega_2)f(\omega_3)
             +f(\omega_2)}{\omega_1-\omega_2+\omega_3}\nn\\
&&\quad+ \left.
        \frac{f(\omega_1)f(\omega_2)
             +f(\omega_1)f(\omega_3)
             -f(\omega_2)f(\omega_3)
             +f(\omega_1)}{-\omega_1+\omega_2+\omega_3}\right]\nn \\
&\times& \rho_1(\omega_1,q_1) \rho_2(\omega_2,q_2)\rho_3(\omega_3,q_3)
\,\,. \label{sunset}
\eea
This integral can be split in
an ultraviolet-finite temperature-dependent part,
\bea 
{\cal S}^T&=&\frac{4}{(2\pi)^7}
\int_{0}^{\infty} d \omega_1\,d \omega_2\,d \omega_3
\, dq_1\, q_1\, dq_2\, q_2\,dq_3\, q_3\,
\Theta\left(|q_1-q_2|\leq q_3\leq q_1+q_2\right)\nn\\
&\times&\left[\frac{f(\omega_1)f(\omega_2)+f(\omega_1)f(\omega_3)
          +f(\omega_3)f(\omega_2)}{
            \omega_1+\omega_2+\omega_3}\right.\nn\\
&&\quad+\frac{f(\omega_1)f(\omega_3)
             -f(\omega_1)f(\omega_2)
             +f(\omega_2)f(\omega_3)}{
            \omega_1+\omega_2-\omega_3}\nn\\
&&\quad+\frac{f(\omega_1)f(\omega_2)
             -f(\omega_1)f(\omega_3)
             +f(\omega_2)f(\omega_3)}{
            \omega_1-\omega_2+\omega_3}\nn\\
&&\quad+ \left.
        \frac{f(\omega_1)f(\omega_2)
             +f(\omega_1)f(\omega_3)
             -f(\omega_2)f(\omega_3)}{
-\omega_1+\omega_2+\omega_3}\right]\nn \\
& \times & \rho_1(\omega_1,q_1) \rho_2(\omega_2,q_2)\rho_3(\omega_3,q_3)
\,\,, 
\eea
an ultraviolet-divergent temperature-dependent part, which
is renormalized by introducing a cut-off $\Lambda$ for one of
the $\omega_i$ integrals,
\bea 
{\cal S}^{\Lambda}&=&\frac{4}{(2\pi)^7}
\int_{0}^{\infty} d \omega_1\,d \omega_2\,d \omega_3
\, dq_1\, q_1\, dq_2\, q_2\,dq_3\, q_3\,
\Theta\left(|q_1-q_2|\leq q_3\leq q_1+q_2\right)\nn\\
&\times&\left[\frac{f(\omega_1)\Theta(\Lambda-\omega_2)
          +f(\omega_2)\Theta(\Lambda - \omega_1)
          +f(\omega_3)\Theta(\Lambda - \omega_1)}{
            \omega_1+\omega_2+\omega_3}\right.\nn\\
&&\quad\left.+\frac{f(\omega_3) \Theta(\Lambda - \omega_1)}{
            \omega_1+\omega_2-\omega_3}
       +\frac{f(\omega_2)\Theta(\Lambda - \omega_1)}{
            \omega_1-\omega_2+\omega_3}+ 
        \frac{f(\omega_1)\Theta(\Lambda - \omega_2)}{
-\omega_1+\omega_2+\omega_3}\right]\nn \\
& \times & \rho_1(\omega_1,q_1) \rho_2(\omega_2,q_2)\rho_3(\omega_3,q_3)
\,\,,
\eea
and an ultraviolet-divergent vacuum part, which
is renormalized by introducing a cut-off $\Lambda$ for two of
the $\omega_i$ integrals,
\bea\label{sunset_lambda_lambda}
{\cal S}^{\Lambda\Lambda}&=&\frac{4}{(2\pi)^7}
\int_{0}^{\infty} d \omega_1\,d \omega_2\,d \omega_3
\, dq_1\, q_1\, dq_2\, q_2\,dq_3\, q_3\,
\Theta\left(|q_1-q_2|\leq q_3\leq q_1+q_2\right)\nn\\
&\times&\frac{\Theta(\Lambda - \omega_1)\Theta(\Lambda-\omega_2) }{
            \omega_1+\omega_2+\omega_3}
\rho_1(\omega_1,q_1) \rho_2(\omega_2,q_2)\rho_3(\omega_3,q_3)
\,\,,
\eea
where 
${\cal S}\equiv{\cal S}^{T}+{\cal S}^{\Lambda}+{\cal S}^{\Lambda\Lambda}$.

\section{The renormalized parameters} \label{AppII}

In this appendix, we discuss how to determine the parameters
$\mu^2, \lambda,$ and $H$ of the $O(N)$ model.
At $T=0$, the equations for the $\sigma$-meson and the pion mass
are, cf.\ Eq.\ (\ref{effmass}),
\bse
\bea
M_\sigma^2& = & \mu^2+\frac{4\lambda}{N}
[3f_\pi^2+3{\cal T}_\sigma^\Lambda+(N-1){\cal T}_\pi^\Lambda]\,\,, \\
M_\pi^2& = & \mu^2+\frac{4\lambda}{N}
[f_\pi^2+{\cal T}_\sigma^\Lambda+(N+1){\cal T}_\pi^\Lambda]\,\,,
\eea
\ese
where the zero-temperature contributions of the tadpole
diagrams, ${\cal T}_{\sigma,\pi}^\Lambda$, are defined in Eq.\ 
(\ref{tadpole_lambda}).
Solving these equations for $\lambda$ and $\mu^2$ one obtains
\bse
\bea
\lambda& \equiv& \lambda(\Lambda)=\frac{N}{8}\frac{M_\sigma^2-M_\pi^2}{f_\pi^2
+{\cal T}_\sigma^\Lambda-{\cal T}_\pi^\Lambda}\,\,, \label{lambda_vac}\\
\mu^2& =& -\frac{M_\sigma^2-3 M_\pi^2}{2}-
\frac{4\lambda}{N}(N+2){\cal T}_\pi^\Lambda\;.
\eea
Equation (\ref{condensate})
for the zero-temperature condensate, $\sigma\equiv f_\pi$, can be written as
\be \label{H_vac}
H=f_\pi\left\{M_\sigma^2-\frac{8\lambda f_\pi^2}{N}
+\left(\frac{4\lambda}{N}\right)^2
\left[2(N-1)\,{\cal S}_{\sigma\sigma\sigma}^{\Lambda\Lambda}
+3!\,{\cal S}_{\sigma\pi\pi}^{\Lambda\Lambda}\right]\right\}\;,
\ee
\ese
where ${\cal S}_{\sigma\pi\pi}^{\Lambda\Lambda}$ is
the zero-temperature contribution of the sunset diagram
consisting of two pions and one $\sigma$-meson, and 
${\cal S}_{\sigma\sigma\sigma}^{\Lambda\Lambda}$ is the corresponding one
consisting of three 
$\sigma$-mesons, respectively. These quantities are defined in 
Eq.\ (\ref{sunset_lambda_lambda}).
Since these contributions are small, we neglect them. 

The term $\sim H$ in
the tree-level potential breaks the residual $O(N-1)$ symmetry
of the vacuum explicitly and favors a certain value for
the scalar condensate $\sigma\equiv f_\pi$. 
The sign of this value is determined
by the sign of $H$. For $H>0$, we must have $\sigma >0$.
In order to preserve this relation, we read off from Eq.\ (\ref{H_vac})
that $\lambda$ is restricted from above by $N M_\sigma^2/(8 f_\pi^2)$.
From Eq.\ (\ref{lambda_vac}), this constraint 
leads to a restriction on the possible
values of the cut-off $\Lambda$.
For a given mass of the $\sigma$-meson, the
maximum value for the cut-off, $\Lambda_{\rm max}$, decreases
with increasing $M_\sigma$. For instance, for $M_\sigma = 
400,\, 600,\, 800$ MeV, one obtains $\Lambda_{\rm max} = 340,\, 260,\, 220$
MeV, respectively.
A similar behavior for the $\sigma$-meson mass as function of
the cut-off has been discussed in Ref.\ \cite{Cooper:1994ji}.

\bibliography{paper}

\begin{thebibliography}{47}
\expandafter\ifx\csname natexlab\endcsname\relax\def\natexlab#1{#1}\fi
\expandafter\ifx\csname bibnamefont\endcsname\relax
  \def\bibnamefont#1{#1}\fi
\expandafter\ifx\csname bibfnamefont\endcsname\relax
  \def\bibfnamefont#1{#1}\fi
\expandafter\ifx\csname citenamefont\endcsname\relax
  \def\citenamefont#1{#1}\fi
\expandafter\ifx\csname url\endcsname\relax
  \def\url#1{\texttt{#1}}\fi
\expandafter\ifx\csname urlprefix\endcsname\relax\def\urlprefix{URL }\fi
\providecommand{\bibinfo}[2]{#2}
\providecommand{\eprint}[2][]{\url{#2}}

\bibitem[{\citenamefont{'t~Hooft}(1986)}]{'tHooft:1986nc}
\bibinfo{author}{\bibfnamefont{G.}~\bibnamefont{'t~Hooft}},
  \bibinfo{journal}{Phys. Rept.} \textbf{\bibinfo{volume}{142}},
  \bibinfo{pages}{357} (\bibinfo{year}{1986}).

\bibitem[{\citenamefont{Vafa and Witten}(1984)}]{Vafa:1984tf}
\bibinfo{author}{\bibfnamefont{C.}~\bibnamefont{Vafa}} \bibnamefont{and}
  \bibinfo{author}{\bibfnamefont{E.}~\bibnamefont{Witten}},
  \bibinfo{journal}{Nucl. Phys.} \textbf{\bibinfo{volume}{B234}},
  \bibinfo{pages}{173} (\bibinfo{year}{1984}).

\bibitem[{\citenamefont{Karsch}(2002)}]{Karsch:2001cy}
\bibinfo{author}{\bibfnamefont{F.}~\bibnamefont{Karsch}},
  \bibinfo{journal}{Lect. Notes Phys.} \textbf{\bibinfo{volume}{583}},
  \bibinfo{pages}{209} (\bibinfo{year}{2002}),
  \eprint[http://arXiv.org/abs]{hep-lat/0106019}.

\bibitem[{\citenamefont{Laermann and Philipsen}(2003)}]{Laermann:2003cv}
\bibinfo{author}{\bibfnamefont{E.}~\bibnamefont{Laermann}} \bibnamefont{and}
  \bibinfo{author}{\bibfnamefont{O.}~\bibnamefont{Philipsen}},
  \bibinfo{journal}{Ann. Rev. Nucl. Part. Sci.} \textbf{\bibinfo{volume}{53}},
  \bibinfo{pages}{163} (\bibinfo{year}{2003}), \eprint{hep-ph/0303042}.

\bibitem[{\citenamefont{Fodor and Katz}(2004)}]{Fodor:2004nz}
\bibinfo{author}{\bibfnamefont{Z.}~\bibnamefont{Fodor}} \bibnamefont{and}
  \bibinfo{author}{\bibfnamefont{S.~D.} \bibnamefont{Katz}},
  \bibinfo{journal}{JHEP} \textbf{\bibinfo{volume}{04}}, \bibinfo{pages}{050}
  (\bibinfo{year}{2004}), \eprint{hep-lat/0402006}.

\bibitem[{\citenamefont{de~Forcrand and Philipsen}(2002)}]{deForcrand:2002ci}
\bibinfo{author}{\bibfnamefont{P.}~\bibnamefont{de~Forcrand}} \bibnamefont{and}
  \bibinfo{author}{\bibfnamefont{O.}~\bibnamefont{Philipsen}},
  \bibinfo{journal}{Nucl. Phys.} \textbf{\bibinfo{volume}{B642}},
  \bibinfo{pages}{290} (\bibinfo{year}{2002}), \eprint{hep-lat/0205016}.

\bibitem[{\citenamefont{Levy}(1967)}]{Levy}
\bibinfo{author}{\bibfnamefont{M.}~\bibnamefont{Levy}}, \bibinfo{journal}{Nuovo
  Cim.} \textbf{\bibinfo{volume}{52}}, \bibinfo{pages}{23}
  (\bibinfo{year}{1967}).

\bibitem[{\citenamefont{Gell-Mann and Levy}(1960)}]{Gell-Mann:1960np}
\bibinfo{author}{\bibfnamefont{M.}~\bibnamefont{Gell-Mann}} \bibnamefont{and}
  \bibinfo{author}{\bibfnamefont{M.}~\bibnamefont{Levy}},
  \bibinfo{journal}{Nuovo Cim.} \textbf{\bibinfo{volume}{16}},
  \bibinfo{pages}{705} (\bibinfo{year}{1960}).

\bibitem[{\citenamefont{Dolan and Jackiw}(1974)}]{Dolan:1974qd}
\bibinfo{author}{\bibfnamefont{L.}~\bibnamefont{Dolan}} \bibnamefont{and}
  \bibinfo{author}{\bibfnamefont{R.}~\bibnamefont{Jackiw}},
  \bibinfo{journal}{Phys. Rev.} \textbf{\bibinfo{volume}{D9}},
  \bibinfo{pages}{3320} (\bibinfo{year}{1974}).

\bibitem[{\citenamefont{Braaten and
  Pisarski}(1990{\natexlab{a}})}]{Braaten:1989kk}
\bibinfo{author}{\bibfnamefont{E.}~\bibnamefont{Braaten}} \bibnamefont{and}
  \bibinfo{author}{\bibfnamefont{R.~D.} \bibnamefont{Pisarski}},
  \bibinfo{journal}{Phys. Rev. Lett.} \textbf{\bibinfo{volume}{64}},
  \bibinfo{pages}{1338} (\bibinfo{year}{1990}{\natexlab{a}}).

\bibitem[{\citenamefont{Braaten and
  Pisarski}(1990{\natexlab{b}})}]{Braaten:1989mz}
\bibinfo{author}{\bibfnamefont{E.}~\bibnamefont{Braaten}} \bibnamefont{and}
  \bibinfo{author}{\bibfnamefont{R.~D.} \bibnamefont{Pisarski}},
  \bibinfo{journal}{Nucl. Phys.} \textbf{\bibinfo{volume}{B337}},
  \bibinfo{pages}{569} (\bibinfo{year}{1990}{\natexlab{b}}).

\bibitem[{\citenamefont{Cornwall et~al.}(1974)\citenamefont{Cornwall, Jackiw,
  and Tomboulis}}]{Cornwall:1974vz}
\bibinfo{author}{\bibfnamefont{J.~M.} \bibnamefont{Cornwall}},
  \bibinfo{author}{\bibfnamefont{R.}~\bibnamefont{Jackiw}}, \bibnamefont{and}
  \bibinfo{author}{\bibfnamefont{E.}~\bibnamefont{Tomboulis}},
  \bibinfo{journal}{Phys. Rev.} \textbf{\bibinfo{volume}{D10}},
  \bibinfo{pages}{2428} (\bibinfo{year}{1974}).

\bibitem[{\citenamefont{Luttinger and Ward}(1960)}]{Luttinger:1960ua}
\bibinfo{author}{\bibfnamefont{J.~M.} \bibnamefont{Luttinger}}
  \bibnamefont{and} \bibinfo{author}{\bibfnamefont{J.~C.} \bibnamefont{Ward}},
  \bibinfo{journal}{Phys. Rev.} \textbf{\bibinfo{volume}{118}},
  \bibinfo{pages}{1417} (\bibinfo{year}{1960}).

\bibitem[{\citenamefont{Baym}(1962)}]{Baym:1962sx}
\bibinfo{author}{\bibfnamefont{G.}~\bibnamefont{Baym}}, \bibinfo{journal}{Phys.
  Rev.} \textbf{\bibinfo{volume}{127}}, \bibinfo{pages}{1391}
  (\bibinfo{year}{1962}).

\bibitem[{\citenamefont{Norton and Cornwall}(1975)}]{Norton:1974bm}
\bibinfo{author}{\bibfnamefont{R.~E.} \bibnamefont{Norton}} \bibnamefont{and}
  \bibinfo{author}{\bibfnamefont{J.~M.} \bibnamefont{Cornwall}},
  \bibinfo{journal}{Ann. Phys.} \textbf{\bibinfo{volume}{91}},
  \bibinfo{pages}{106} (\bibinfo{year}{1975}).

\bibitem[{\citenamefont{Kleinert}(1982)}]{Kleinert:1982ki}
\bibinfo{author}{\bibfnamefont{H.}~\bibnamefont{Kleinert}},
  \bibinfo{journal}{Fortsch. Phys.} \textbf{\bibinfo{volume}{30}},
  \bibinfo{pages}{187} (\bibinfo{year}{1982}).

\bibitem[{\citenamefont{Carrington}(2004)}]{Carrington:2004sn}
\bibinfo{author}{\bibfnamefont{M.~E.} \bibnamefont{Carrington}},
  \bibinfo{journal}{Eur. Phys. J.} \textbf{\bibinfo{volume}{C35}},
  \bibinfo{pages}{383} (\bibinfo{year}{2004}), \eprint{hep-ph/0401123}.

\bibitem[{\citenamefont{Berges}(2004)}]{Berges:2004pu}
\bibinfo{author}{\bibfnamefont{J.}~\bibnamefont{Berges}},
  \bibinfo{journal}{Phys. Rev.} \textbf{\bibinfo{volume}{D70}},
  \bibinfo{pages}{105010} (\bibinfo{year}{2004}), \eprint{hep-ph/0401172}.

\bibitem[{\citenamefont{van Hees and
  Knoll}(2002{\natexlab{a}})}]{vanHees:2002bv}
\bibinfo{author}{\bibfnamefont{H.}~\bibnamefont{van Hees}} \bibnamefont{and}
  \bibinfo{author}{\bibfnamefont{J.}~\bibnamefont{Knoll}},
  \bibinfo{journal}{Phys. Rev.} \textbf{\bibinfo{volume}{D66}},
  \bibinfo{pages}{025028} (\bibinfo{year}{2002}{\natexlab{a}}),
  \eprint[http://arXiv.org/abs]{hep-ph/0203008}.

\bibitem[{\citenamefont{Petropoulos}(1999)}]{Petropoulos:1998gt}
\bibinfo{author}{\bibfnamefont{N.}~\bibnamefont{Petropoulos}},
  \bibinfo{journal}{J. Phys.} \textbf{\bibinfo{volume}{G25}},
  \bibinfo{pages}{2225} (\bibinfo{year}{1999}),
  \eprint[http://arXiv.org/abs]{hep-ph/9807331}.

\bibitem[{\citenamefont{Lenaghan and Rischke}(2000)}]{Lenaghan:1999si}
\bibinfo{author}{\bibfnamefont{J.~T.} \bibnamefont{Lenaghan}} \bibnamefont{and}
  \bibinfo{author}{\bibfnamefont{D.~H.} \bibnamefont{Rischke}},
  \bibinfo{journal}{J. Phys.} \textbf{\bibinfo{volume}{G26}},
  \bibinfo{pages}{431} (\bibinfo{year}{2000}),
  \eprint[http://arXiv.org/abs]{nucl-th/9901049}.

\bibitem[{\citenamefont{Lenaghan et~al.}(2000)\citenamefont{Lenaghan, Rischke,
  and Schaffner-Bielich}}]{Lenaghan:2000ey}
\bibinfo{author}{\bibfnamefont{J.~T.} \bibnamefont{Lenaghan}},
  \bibinfo{author}{\bibfnamefont{D.~H.} \bibnamefont{Rischke}},
  \bibnamefont{and}
  \bibinfo{author}{\bibfnamefont{J.}~\bibnamefont{Schaffner-Bielich}},
  \bibinfo{journal}{Phys. Rev.} \textbf{\bibinfo{volume}{D62}},
  \bibinfo{pages}{085008} (\bibinfo{year}{2000}),
  \eprint[http://arXiv.org/abs]{nucl-th/0004006}.

\bibitem[{\citenamefont{R{\"o}der et~al.}(2003)\citenamefont{R{\"o}der,
  Ruppert, and Rischke}}]{Roder:2003uz}
\bibinfo{author}{\bibfnamefont{D.}~\bibnamefont{R{\"o}der}},
  \bibinfo{author}{\bibfnamefont{J.}~\bibnamefont{Ruppert}}, \bibnamefont{and}
  \bibinfo{author}{\bibfnamefont{D.~H.} \bibnamefont{Rischke}},
  \bibinfo{journal}{Phys. Rev.} \textbf{\bibinfo{volume}{D68}},
  \bibinfo{pages}{016003} (\bibinfo{year}{2003}), \eprint{nucl-th/0301085}.

\bibitem[{\citenamefont{Baym and Grinstein}(1977)}]{Baym:1977qb}
\bibinfo{author}{\bibfnamefont{G.}~\bibnamefont{Baym}} \bibnamefont{and}
  \bibinfo{author}{\bibfnamefont{G.}~\bibnamefont{Grinstein}},
  \bibinfo{journal}{Phys. Rev.} \textbf{\bibinfo{volume}{D15}},
  \bibinfo{pages}{2897} (\bibinfo{year}{1977}).

\bibitem[{\citenamefont{Bochkarev and Kapusta}(1996)}]{Bochkarev:1995gi}
\bibinfo{author}{\bibfnamefont{A.}~\bibnamefont{Bochkarev}} \bibnamefont{and}
  \bibinfo{author}{\bibfnamefont{J.~I.} \bibnamefont{Kapusta}},
  \bibinfo{journal}{Phys. Rev.} \textbf{\bibinfo{volume}{D54}},
  \bibinfo{pages}{4066} (\bibinfo{year}{1996}), \eprint{hep-ph/9602405}.

\bibitem[{\citenamefont{Roh and Matsui}(1998)}]{Roh:1996ek}
\bibinfo{author}{\bibfnamefont{H.-S.} \bibnamefont{Roh}} \bibnamefont{and}
  \bibinfo{author}{\bibfnamefont{T.}~\bibnamefont{Matsui}},
  \bibinfo{journal}{Eur. Phys. J.} \textbf{\bibinfo{volume}{A1}},
  \bibinfo{pages}{205} (\bibinfo{year}{1998}), \eprint{nucl-th/9611050}.

\bibitem[{\citenamefont{Amelino-Camelia}(1997)}]{Amelino-Camelia:1997dd}
\bibinfo{author}{\bibfnamefont{G.}~\bibnamefont{Amelino-Camelia}},
  \bibinfo{journal}{Phys. Lett.} \textbf{\bibinfo{volume}{B407}},
  \bibinfo{pages}{268} (\bibinfo{year}{1997}), \eprint{hep-ph/9702403}.

\bibitem[{\citenamefont{Pisarski and Wilczek}(1984)}]{Pisarski:1984ms}
\bibinfo{author}{\bibfnamefont{R.~D.} \bibnamefont{Pisarski}} \bibnamefont{and}
  \bibinfo{author}{\bibfnamefont{F.}~\bibnamefont{Wilczek}},
  \bibinfo{journal}{Phys. Rev.} \textbf{\bibinfo{volume}{D29}},
  \bibinfo{pages}{338} (\bibinfo{year}{1984}).

\bibitem[{\citenamefont{Aouissat et~al.}(1998)\citenamefont{Aouissat, Bohr, and
  Wambach}}]{Aouissat:1997nu}
\bibinfo{author}{\bibfnamefont{Z.}~\bibnamefont{Aouissat}},
  \bibinfo{author}{\bibfnamefont{O.}~\bibnamefont{Bohr}}, \bibnamefont{and}
  \bibinfo{author}{\bibfnamefont{J.}~\bibnamefont{Wambach}},
  \bibinfo{journal}{Mod. Phys. Lett.} \textbf{\bibinfo{volume}{A13}},
  \bibinfo{pages}{1827} (\bibinfo{year}{1998}), \eprint{hep-ph/9710419}.

\bibitem[{\citenamefont{van Hees and
  Knoll}(2002{\natexlab{b}})}]{vanHees:2001ik}
\bibinfo{author}{\bibfnamefont{H.}~\bibnamefont{van Hees}} \bibnamefont{and}
  \bibinfo{author}{\bibfnamefont{J.}~\bibnamefont{Knoll}},
  \bibinfo{journal}{Phys. Rev.} \textbf{\bibinfo{volume}{D65}},
  \bibinfo{pages}{025010} (\bibinfo{year}{2002}{\natexlab{b}}),
  \eprint[http://arXiv.org/abs]{hep-ph/0107200}.

\bibitem[{\citenamefont{Van~Hees and Knoll}(2002)}]{VanHees:2001pf}
\bibinfo{author}{\bibfnamefont{H.}~\bibnamefont{Van~Hees}} \bibnamefont{and}
  \bibinfo{author}{\bibfnamefont{J.}~\bibnamefont{Knoll}},
  \bibinfo{journal}{Phys. Rev.} \textbf{\bibinfo{volume}{D65}},
  \bibinfo{pages}{105005} (\bibinfo{year}{2002}),
  \eprint[http://arXiv.org/abs]{hep-ph/0111193}.

\bibitem[{\citenamefont{Aarts et~al.}(2002)\citenamefont{Aarts, Ahrensmeier,
  Baier, Berges, and Serreau}}]{Aarts:2002dj}
\bibinfo{author}{\bibfnamefont{G.}~\bibnamefont{Aarts}},
  \bibinfo{author}{\bibfnamefont{D.}~\bibnamefont{Ahrensmeier}},
  \bibinfo{author}{\bibfnamefont{R.}~\bibnamefont{Baier}},
  \bibinfo{author}{\bibfnamefont{J.}~\bibnamefont{Berges}}, \bibnamefont{and}
  \bibinfo{author}{\bibfnamefont{J.}~\bibnamefont{Serreau}},
  \bibinfo{journal}{Phys. Rev.} \textbf{\bibinfo{volume}{D66}},
  \bibinfo{pages}{045008} (\bibinfo{year}{2002}), \eprint{hep-ph/0201308}.

\bibitem[{\citenamefont{Verschelde and Coppens}(1992)}]{Verschelde:1992bs}
\bibinfo{author}{\bibfnamefont{H.}~\bibnamefont{Verschelde}} \bibnamefont{and}
  \bibinfo{author}{\bibfnamefont{M.}~\bibnamefont{Coppens}},
  \bibinfo{journal}{Phys. Lett.} \textbf{\bibinfo{volume}{B287}},
  \bibinfo{pages}{133} (\bibinfo{year}{1992}).

\bibitem[{\citenamefont{Verschelde}(2001)}]{Verschelde:2000dz}
\bibinfo{author}{\bibfnamefont{H.}~\bibnamefont{Verschelde}},
  \bibinfo{journal}{Phys. Lett.} \textbf{\bibinfo{volume}{B497}},
  \bibinfo{pages}{165} (\bibinfo{year}{2001}), \eprint{hep-th/0009123}.

\bibitem[{\citenamefont{Baacke and Michalski}(2003)}]{Baacke:2002pi}
\bibinfo{author}{\bibfnamefont{J.}~\bibnamefont{Baacke}} \bibnamefont{and}
  \bibinfo{author}{\bibfnamefont{S.}~\bibnamefont{Michalski}},
  \bibinfo{journal}{Phys. Rev.} \textbf{\bibinfo{volume}{D67}},
  \bibinfo{pages}{085006} (\bibinfo{year}{2003}), \eprint{hep-ph/0210060}.

\bibitem[{\citenamefont{{Eidelman} et~al.}(2004)\citenamefont{{Eidelman},
  {Hayes}, {Olive}, {Aguilar-Benitez}, {Amsler}, {Asner}, {Babu}, {Barnett},
  {Beringer}, {Burchat} et~al.}}]{PDBook}
\bibinfo{author}{\bibfnamefont{S.}~\bibnamefont{{Eidelman}}},
  \bibinfo{author}{\bibfnamefont{K.}~\bibnamefont{{Hayes}}},
  \bibinfo{author}{\bibfnamefont{K.}~\bibnamefont{{Olive}}},
  \bibinfo{author}{\bibfnamefont{M.}~\bibnamefont{{Aguilar-Benitez}}},
  \bibinfo{author}{\bibfnamefont{C.}~\bibnamefont{{Amsler}}},
  \bibinfo{author}{\bibfnamefont{D.}~\bibnamefont{{Asner}}},
  \bibinfo{author}{\bibfnamefont{K.}~\bibnamefont{{Babu}}},
  \bibinfo{author}{\bibfnamefont{R.}~\bibnamefont{{Barnett}}},
  \bibinfo{author}{\bibfnamefont{J.}~\bibnamefont{{Beringer}}},
  \bibinfo{author}{\bibfnamefont{P.}~\bibnamefont{{Burchat}}},
  \bibnamefont{et~al.}, \bibinfo{journal}{{Physics Letters B}}
  \textbf{\bibinfo{volume}{592}}, \bibinfo{pages}{1+} (\bibinfo{year}{2004}),
  \urlprefix\url{http://pdg.lbl.gov}.

\bibitem[{\citenamefont{R{\"o}der}(2005)}]{Roder:2005qy}
\bibinfo{author}{\bibfnamefont{D.}~\bibnamefont{R{\"o}der}}
  (\bibinfo{year}{2005}), \eprint{hep-ph/0509232}.

\bibitem[{\citenamefont{Le~Bellac}(2000)}]{leBellac}
\bibinfo{author}{\bibfnamefont{M.}~\bibnamefont{Le~Bellac}},
  \emph{\bibinfo{title}{Thermal Field Theory}} (\bibinfo{publisher}{Cambridge
  University Press}, \bibinfo{year}{2000}).

\bibitem[{\citenamefont{Weldon}(1983)}]{Weldon:1983jn}
\bibinfo{author}{\bibfnamefont{H.~A.} \bibnamefont{Weldon}},
  \bibinfo{journal}{Phys. Rev.} \textbf{\bibinfo{volume}{D28}},
  \bibinfo{pages}{2007} (\bibinfo{year}{1983}).

\bibitem[{\citenamefont{Beckmann et~al.}()\citenamefont{Beckmann, Wilms, and
  Rischke}}]{beckmannphd}
\bibinfo{author}{\bibfnamefont{C.}~\bibnamefont{Beckmann}},
  \bibinfo{author}{\bibfnamefont{S.}~\bibnamefont{Wilms}}, \bibnamefont{and}
  \bibinfo{author}{\bibfnamefont{D.~H.} \bibnamefont{Rischke}},
  \bibinfo{note}{(in preparation)}.

\bibitem[{\citenamefont{van Hees and Knoll}(2000)}]{vanHees:2000bp}
\bibinfo{author}{\bibfnamefont{H.}~\bibnamefont{van Hees}} \bibnamefont{and}
  \bibinfo{author}{\bibfnamefont{J.}~\bibnamefont{Knoll}},
  \bibinfo{journal}{Nucl. Phys.} \textbf{\bibinfo{volume}{A683}},
  \bibinfo{pages}{369} (\bibinfo{year}{2000}),
  \eprint[http://arXiv.org/abs]{hep-ph/0007070}.

\bibitem[{\citenamefont{Ruppert and Renk}(2005)}]{Ruppert:2004yg}
\bibinfo{author}{\bibfnamefont{J.}~\bibnamefont{Ruppert}} \bibnamefont{and}
  \bibinfo{author}{\bibfnamefont{T.}~\bibnamefont{Renk}},
  \bibinfo{journal}{Phys. Rev.} \textbf{\bibinfo{volume}{C71}},
  \bibinfo{pages}{064903} (\bibinfo{year}{2005}), \eprint{nucl-th/0412047}.

\bibitem[{\citenamefont{Str{\"u}ber and Rischke}()}]{strueberdipl}
\bibinfo{author}{\bibfnamefont{S.}~\bibnamefont{Str{\"u}ber}} \bibnamefont{and}
  \bibinfo{author}{\bibfnamefont{D.~H.} \bibnamefont{Rischke}},
  \bibinfo{note}{(in preparation)}.

\bibitem[{\citenamefont{Rapp and Wambach}(2000)}]{Rapp:1999ej}
\bibinfo{author}{\bibfnamefont{R.}~\bibnamefont{Rapp}} \bibnamefont{and}
  \bibinfo{author}{\bibfnamefont{J.}~\bibnamefont{Wambach}},
  \bibinfo{journal}{Adv. Nucl. Phys.} \textbf{\bibinfo{volume}{25}},
  \bibinfo{pages}{1} (\bibinfo{year}{2000}), \eprint{hep-ph/9909229}.

\bibitem[{\citenamefont{Das}(1997)}]{das}
\bibinfo{author}{\bibfnamefont{A.}~\bibnamefont{Das}},
  \emph{\bibinfo{title}{Finite Temperature Field Theory}}
  (\bibinfo{publisher}{World Scientific}, \bibinfo{year}{1997}).

\bibitem[{\citenamefont{Kapusta}(1993)}]{kapusta}
\bibinfo{author}{\bibfnamefont{J.~I.} \bibnamefont{Kapusta}},
  \emph{\bibinfo{title}{Finite Temperature Field Theory}}
  (\bibinfo{publisher}{Cambridge University Press}, \bibinfo{year}{1993}).

\bibitem[{\citenamefont{Cooper et~al.}(1995)\citenamefont{Cooper, Kluger,
  Mottola, and Paz}}]{Cooper:1994ji}
\bibinfo{author}{\bibfnamefont{F.}~\bibnamefont{Cooper}},
  \bibinfo{author}{\bibfnamefont{Y.}~\bibnamefont{Kluger}},
  \bibinfo{author}{\bibfnamefont{E.}~\bibnamefont{Mottola}}, \bibnamefont{and}
  \bibinfo{author}{\bibfnamefont{J.~P.} \bibnamefont{Paz}},
  \bibinfo{journal}{Phys. Rev.} \textbf{\bibinfo{volume}{D51}},
  \bibinfo{pages}{2377} (\bibinfo{year}{1995}), \eprint{hep-ph/9404357}.

\end{thebibliography}
\newpage

\begin{figure}
\includegraphics[height=2cm]{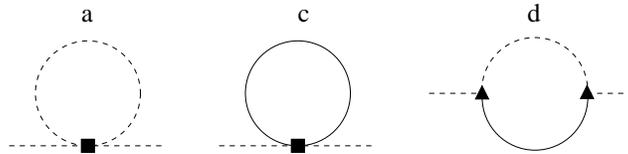}
\caption{The self-energy of the pion. The diagrams a and c
are tadpole contributions generated by cutting an internal pion line in the
double-bubble diagrams a and c in Fig.\ \ref{paper1}. The diagram 
d is the one-loop contribution generated by cutting an internal pion
line in the sunset diagram d in Fig.~\ref{paper1}. 
Lines and vertices as in Fig.\ \ref{paper1}.}
\label{selfenergy_pion}
\end{figure}

\end{document}